# Toeplitz Inverse Eigenvalue Problem (ToIEP) and Random Matrix Theory (RMT) Support for the Toeplitz Covariance Matrix Estimation

Yuri Abramovich, *Life Fellow, IEEE*, Tanit Pongsiri

***Abstract*****—"Toeplitzification" or "redundancy (spatial) averaging", the well-known routine for deriving the Toeplitz covariance matrix estimate from the standard sample covariance matrix, recently regained new attention due to the important Random Matrix Theory (RMT) findings. The asymptotic consistency in the spectral norm was proven for the Kolmogorov's asymptotics when the matrix dimension $N$ and independent identically distributed (i.i.d.) sample volume $T$ both tended to infinity ($N \to \infty, T \to \infty, T/N \to c > 0$). These novel RMT results encouraged us to reassess the well-known drawback of the redundancy averaging methodology, which is the generation of the negative minimal eigenvalues for covariance matrices with big eigenvalues spread, typical for most covariance matrices of interest. We demonstrate that for this type of Toeplitz covariance matrices, convergence in the spectral norm does not prevent the generation of negative eigenvalues, even for the sample volume $T$ that significantly exceeds the covariance matrix dimension ($T \gg$ N). We demonstrate that the ad-hoc attempts to remove the negative eigenvalues by the proper diagonal loading result in solutions with the very low likelihood. We demonstrate that attempts to exploit Newton's type iterative algorithms, designed to produce a Hermitian Toeplitz matrix with the given eigenvalues lead to the very poor likelihood of the very slowly converging solution to the desired eigenvalues. Finally, we demonstrate that the proposed algorithm for restoration of a positive definite (p.d.) Hermitian Toeplitz matrix with the specified Maximum Entropy spectrum, allows for the transformation of the (unstructured) Hermitian maximum likelihood (ML) sample matrix estimate in a p.d. Toeplitz matrix with sufficiently high likelihood.**

## I. Introduction

Estimation of the $N$-variate Hermitian Toeplitz covariance matrix, given as a set of $T$ i.i.d Gaussian training samples $\mathbf{X}_t \in \mathbb{C}^{N\,x\,1}, t = 1, \ldots T$, is one of the classical signal processing problems introduced in [1], [2]. Since no closed-form maximum likelihood solution has been provided for this problem [3], numerous ad-hoc techniques were developed over the decades [16]-[19]. One of the earliest and simplest techniques that even predates [2] is the so-called "redundancy" or "spatial" averaging, applied to the unstructured ML sample covariance matrix estimate. This procedure averages all redundant estimates of a particular correlation lag in the sample matrix and replaces each of them by their average [4], [5]. The early studies of this technique revealed its ability to generate the negative smallest eigenvalues when the true minimal eigenvalues of the covariance matrix are close to zero. Yet, with the introduction of the Random Matrix Theory methodology in recent years [6], [20], the interest in this sample matrix transformation technique re-emerged. Specifically, in the above mentioned RMT asymptotic regime ($N, T \to \infty$, $T/N \to c$) the consistency in the spectral norm of the redundancy averaged Toeplitz covariance matrix $\mathbf{T}_N$ was proven in [20]:

$$\lim_{T\to\infty} \|\mathbf{T}_N - \mathbf{T}_N^{\mathrm{ra}}\| \overset{as}{\to} 0 \tag{1}$$

Recall, that the spectral norm of a square matrix $\mathbf{A}$ is defined as the square root of the maximal eigenvalue of the product $\mathbf{A}^{\mathrm{H}}\mathbf{A}$, which does not preclude existence of the negative eigenvalues in the redundancy averaged sample matrix $\widehat{\mathbf{T}}_N^{\mathrm{ra}}$. In [6], under the assumption that ($N \to \infty, T \to \infty, T/N \to c > 0$) and that the covariance lags $\mathbf{t}_n$ are summable with $\mathbf{t}_o > 0$, the following Theorem 2 was proven:

Theorem 2. Under some technical assumptions, for the redundancy averaged matrix $\widehat{\mathbf{T}}_N^{\mathrm{ra}}$ and any $x > 0$, we have

$$\begin{aligned} \mathrm{P}[\|[\widehat{\mathbf{T}}_N^{\mathrm{ra}} - \mathbf{T}_N]\| > x] \\ \le \exp\left(-\frac{cTx^2}{4\|\boldsymbol{\gamma}\|_\infty^2 \log T}(1 + O(1))\right) \end{aligned} \tag{2}$$

where $O(1)$ is with respect to $T$, and depends on $x$.

In (2),

$$\begin{gathered} \gamma(\lambda) \triangleq \sum_{k=-\infty}^{\infty} \mathbf{t}_k e^{-ik\lambda}, \quad \lambda \in [0,2\pi] \\ \|\boldsymbol{\Gamma}_N\| \le \|\gamma\|_\infty \end{gathered} \tag{3}$$

The consequence of this theorem proven in [6] is that $\|[\widehat{\mathbf{T}}_N^{\mathrm{ra}} - \mathbf{T}_N]\| \to 0$ almost surely as $\mathbf{T}_N \to \infty$. The slower convergence rate $T/\log T$ in the redundancy averaged matrix is explained in [6] by the inaccuracy of the estimates of $\mathbf{t}_n$ close to $N-1$. In [7], this slower rate $(\log(NT)/NT\,)$^$(2\beta/(2\beta -$

Yuri Abramovich and Tanit Pongsiri are with WR Systems, Ltd., Fairfax, VA 22030, USA. Emails: yabramovich@wrsystems.com, tpongsiri@wrsystems.com

1)) is defined as the optimum (!) convergence rate for a specific sub-class of the Toeplitz matrices. Once again, the convergence of the spectral norm $\|[\widehat{\mathbf{T}}_N - \mathbf{T}_N]\| \to 0$ was considered in Theorem 1 [20], but that does not suggest that this convergence takes place in the class of the p.d. Toeplitz matrices.

Note that this convergence rate in the spectral norm differs from the Cramer-Rao convergence, defined in [12] as $(1/T)\boldsymbol{\mathcal{J}}_{\Sigma}^{-1}(m,n)$, where $\boldsymbol{\mathcal{J}}_{\boldsymbol{\Sigma}}(m,n)$ is the Fisher Information Matrix. Recall that the Cramer-Rao Bound (CRB) specifies the mean square error of the matrix's elements, which is different from the spectral norm. This different convergence rate in different estimation error metrics is somewhat disturbing. In [20], the MUSIC performance improvement with the redundancy averaged matrices was demonstrated for a certain scenario. Yet, in a number of important applications the inverse of the estimated Toeplitz covariance matrix has to be used and the solutions with the negative eigenvalues are completely inappropriate here. It is not difficult to demonstrate that the ad-hoc attempts to get rid of negative eigenvalues by the appropriate diagonal loading result in solutions with very low likelihood.

Therefore our study has two tightly connected goals. First, in the view of the recent RMT theoretical results on asymptotic consistency of the redundancy averaged Toeplitz matrices, to re-examine the negative eigenvalue generation phenomenon and in general, the likelihood of the estimates of the Toeplitz Hermitian covariance matrices with the negative eigenvalues removed. In this experimental study we consider a specific Hermitian Toeplitz covariance matrix, quite typical for some applications. Not surprisingly, this analysis confirmed observations made long ago [4]: for the considered ("spiked") class of the Hermitian Toeplitz covariance matrices negative eigenvalues are generated even for the extremely (!) large sample volume *T*, many times exceeding the matrix dimension *N*. By appropriate diagonal loading one can remove the negative eigenvalues, but the likelihood of the derived solutions is proven to be very low.

In order to improve the redundancy averaged solutions, we employ some iterative Newton's-like techniques in attempt to achieve the Toeplitz matrices with the RMT-modified eigenvalues [11]. This attempt is performed in anticipation that starting from the redundancy averaged solution, these iterations should lead to a close to the optimum solution. This expectation is quite crucial since the Hermitian Toeplitz matrix cannot be uniquely reconstructed given the eigenvalues only. In the famous theorem of Landau [22], it stated that "every set of *n* real numbers is the spectrum of *n x n* real symmetric Toeplitz matrix". For a symmetric real-valued Toeplitz matrix only a limited number of solutions with the same eigenspectrum exists [14], [23]. Therefore, only by starting from a redundancy averaged Hermitian matrix that's close enough, we hoped that we may get the desired eigenspectrum, not deviating far from the optimum solution. Unfortunately, our investigation demonstrated the opposite tendency: while very slowly approaching the specified eigenvalues, we rapidly moved away from the optimum (maximum likelihood) solution. For this reason, the techniques that may satisfactorily work for the real-valued symmetric Toeplitz matrices are found inappropriate for the case of the Hermitian Toeplitz matrices.

Since the unique reconstruction of the Hermitian Toeplitz matrix by its eigenspectrum is impossible, our second goal was to consider different options. Specifically, we employed our method developed in [24] of the unique Hermitian p.d. Toeplitz matrix reconstruction given the Maximum Entropy spectrum, calculated for the sample covariance matrix. We demonstrate that this reconstruction results in a p.d. Hermitian Toeplitz matrices with a significantly higher likelihood ratio, statistically approaching the likelihood ratio value generated by the true covariance matrix.

To reflect this development, in Sec II we experimentally investigate the negative eigenvalues production by redundancy averaging for the particular "spiked" Toeplitz Hermitian covariance matrix. In Sec III we introduce our techniques we used for transformation of the sample Hermitian matrix into the p.d. Toeplitz matrix with high likelihood. In Sec IV we report on the results of the Monte-Carlo simulations. In Sec V we conclude our paper.

## II. Negative Eigenvalues Generation By The Sample Matrix Redundancy Averaging For "Spiked" Covariance Matrices

In most applications, the Toeplitz covariance matrix belongs to the class of the so-called "spiked" covariance matrices. The most typical example of such matrices provides the covariance matrix of the mixture of $k < N$ independently fluctuating plane waves with the directions of arrival $\theta_j, j = 1, \dots, k$ in an *N*-element uniform linear antenna (ULA), when

$$\mathbf{T}_N = \sigma_n^2 \mathbf{I}_N + \mathbf{A}(k)\mathbf{D}_k\mathbf{A}^{\mathrm{H}}(k) \tag{4}$$

$$\mathbf{A}(k) = [a(\theta_1), a(\theta_2), \dots, a(\theta_k)] \in \mathbb{C}^{N\,x\,\mathrm{k}} \tag{5}$$

$$a(\theta_j) = [1, \exp(i\frac{2\pi d}{\lambda}\sin\theta_j), \dots, \exp(i\frac{2\pi d}{\lambda}(N-1)\sin\theta_j)]^{\mathrm{T}} \tag{6}$$

$$\mathbf{D}_k = \mathrm{diag}[\sigma_1^2, \sigma_2^2, \dots, \sigma_k^2], \quad \sigma_j^2 \gg \sigma_n^2 \tag{7}$$

In fact, some of the signal powers $\sigma_j^2$ may be close to the additive noise power while some of the direction of arrivals (DOA) $\theta_j$ may be close to each other, which result in some minimal "signal subspace" eigenvalues $\boldsymbol{\lambda}_1 > \boldsymbol{\lambda}_2 > \cdots > \boldsymbol{\lambda}_k$ approaching the noise power $\sigma_n^2$. Yet the additive noise poiwer $\sigma_n^2 \ll \sigma_j^2$ is smaller than the "signal subspace" eigenvalues.

Another example is provided by the spatial covariance matrix of the clutter returns in HF over-the-horizon radar (HF-OTHR), illuminated by the transmitting antennas with the "finger beam" beampattern:

$$\mathbf{T}_N = \text{sinc}\,(W_1) + 0.5\,\text{diag}\,(\theta_o)\,\text{sinc}\,(W_2)\text{diag}^{\text{H}}(\theta_o), \tag{8}$$

where

$$\text{sinc}\,(W) = [\frac{\sin 2\pi W\,(n-k)}{\pi(n-k)}]_{n,k=1,N}, \quad W \le 0.5 \tag{9}$$

$$\text{diag}\,(\theta_o) = \text{diag}[1, \exp\left(i\frac{2\pi d}{\lambda}\sin\theta_o\right), \dots, \exp\left(i(N-1)\frac{2\pi d}{\lambda}\sin\theta_o\right)] \tag{10}$$

For the Monte Carlo simulations we selected $N = 17, W_1 = 0.2, W_2 = 0.1, \theta_o = 20^{\text{o}}, d/\lambda = 0.5$ and $\sigma_n^2 = 10^{-4}$. The eigenvalues of this matrix are introduced in TABLE I.

TABLE I.

| $\boldsymbol{\lambda}(\boldsymbol{T}_N)$ |
|---|
| 1.49641081 |
| 1.42482983 |
| 1.13675988 |
| 1.00087182 |
| 1.00008334 |
| 0.99237031 |
| 0.81498939 |
| 0.45059824 |
| 0.15840976 |
| 0.02362155 |
| 0.00204107 |
| 0.00020973 |
| 0.00010417 |
| 0.00010011 |
| 0.00010000 |
| 0.00010000 |
| 0.00010000 |

We can see that the last four eigenvalues are accurately equal to the noise power, while $\lambda_{12} = 0.00020973$ exceeds the minimum eigenvalue by 3dB only. Yet, all these last six eigenvalues are ~40dB weaker than the first eigenvalue $\lambda_1 = 1.4964$. As one can see, the sharp boundary between the noise and signal subspace eigenvalues are somewhat blurred for $N = 17$, and therefore, depending on the number of the i.i.d. training samples $\mathbf{X}_t, t = 1, \dots, T$, the information theoretic criteria [8] may define the noise subspace dimension arbitrarily. For this reason, the most appropriate definition of the "spiked" covariance matrices is provided in [21].

Specifically, we consider $\mathbf{T}_N$ to be the spiked covariance matrix, if

$$\boldsymbol{\lambda}_1 \ge \boldsymbol{\lambda}_2 \ge \cdots \ge \boldsymbol{\lambda}_m \ge \boldsymbol{\lambda}_{m+1} \ge \cdots \ge \boldsymbol{\lambda}_N > 0 \tag{11}$$

with the "spiked" eigenvalues being bounded by

$$c_o \le \boldsymbol{\lambda}_j \le C_o, \quad j \le m \tag{12}$$

for the constants $c_o, C_o > 0$, and the spiked eigenvalues are well separated, that is $\exists\, \delta_o > 0$ such that $\min_{j \le m}(\boldsymbol{\lambda}_j - \boldsymbol{\lambda}_{j+1}) \ge \delta_o$. Similar to [21], we do not need to order the leading eigenvalues or require them to diverge. For simplicity, we assume different "signal subspace" eigenvalues and do not require the "noise subspace" eigenvalues to be identical. Under this and some extra technical assumptions, in [21] it was proven that the "signal subspace" sample eigenvalues $\{\hat{\boldsymbol{\lambda}}_j\}_{j=1,\dots,m}$ have independent limiting distribution, and that (Theorem 3.1 in [21])

$$\sqrt{T}\left\{\frac{\hat{\boldsymbol{\lambda}}_j}{\boldsymbol{\lambda}_j} - \left(1 - \bar{c}c_j + O_p\left(\boldsymbol{\lambda}_j^{-1}\sqrt{\frac{N}{T}}\right)\right)\right\} \overset{d}{\Rightarrow} \mathcal{N}(0, k_j - 1) \tag{13}$$

where $k_j$ is the kurtosis of $x_j = \mathbf{T}_N^{-1/2}\mathbf{Y}_j$,

$$c_j = \frac{N}{T\boldsymbol{\lambda}_j}, \quad (N-m)^{-1}\sum_{j=m+1}^{N}\boldsymbol{\lambda}_j = \bar{c} + O(\mathbb{T}^{-1/2}) \tag{14}$$

This theorem reveals the bias in $\hat{\boldsymbol{\lambda}}_j$ is controlled by the terms with the decay rate $N/T\boldsymbol{\lambda}_j$. To get the asymptotically unbiased estimate, it requires

$$c_j = \frac{N}{T\boldsymbol{\lambda}_j} \to 0 \quad \text{for } j \le m \tag{15}$$

In [25] under slightly different assumptions, a similar bound was derived for the spiked eigenvalues:

$$\frac{\hat{\boldsymbol{\lambda}}_j}{\boldsymbol{\lambda}_j} - 1 = O_N\left(\frac{N}{T\boldsymbol{\lambda}_j} + \frac{m^4}{T} + \frac{1}{\boldsymbol{\lambda}_j}\right), \quad \text{for } j = 1, \dots, m \tag{16}$$

Note that if this conditions is applied to the $\lambda_{11} = 0.0002097$ and when it's required that the bias should not lead to the negative domain, we have

$$\frac{N}{T\lambda_{11}} < \lambda_{11}, \quad \text{or} \quad \frac{N}{T(\lambda_{11})^2} < 1 \qquad (17)$$

which suggests that for $N = 17, T > 10^8$ (!) is required to meet this condition. In Theorem 2.5 in [25], under a number of technical conditions, the following bound for the "non-spiked" ("noise" subspace) eigenvalues $\boldsymbol{\lambda}_j (1 \le j - m \le \log T)$ was derived:

$$|\boldsymbol{\lambda}_j - \upsilon_{j-m}| < T^{-\frac{2}{3}-\varepsilon} \qquad (18)$$

In particular, the greatest "non-spiked" eigenvalue $\boldsymbol{\lambda}_{m+1}$ has the limiting Type 1 Tracy-Wisdom distribution.

In [25], $\upsilon_j$ is the $j$-th largest eigenvalue of the matrix $(1/T)\sum_{t=1}^{T} \boldsymbol{\xi}_t \boldsymbol{\Sigma}_1 \boldsymbol{\xi}_t^{\mathrm{H}}$, where $\boldsymbol{\Sigma}_1 = \boldsymbol{\mathcal{U}}_2 \boldsymbol{\Lambda}_{N-m} \boldsymbol{\mathcal{U}}_2^{\mathrm{H}}$, $\boldsymbol{\Lambda}_{N-m}$ is the diagonal matrix of the "nonspiked" (noise subspace) eigenvalues and $\boldsymbol{\mathcal{U}}_2$ is the set of $(N-m)$ noise subspace eigenvalues.

$$\boldsymbol{U} = [\boldsymbol{\mathcal{U}}_1, \boldsymbol{\mathcal{U}}_2]; \quad \mathbf{Y}_t = \mathbf{T}_N^{1/2} \boldsymbol{\xi}_t \qquad (19)$$

This brief discussion of the most recent RMT results on eigenvalues of the sample matrix for the practically relevant "spiked" Toeplitz covariance matrix demonstrates significant differences in asymptotic behavior of the spiked (signal subspace) and the "non-spiked" (noise subspace) eigenvalues of a sample matrix. Obviously, there is no reason to expect that the redundancy averaging applied to the sample matrix should remove this property, allowing to associate the negative eigenvalue generation with the proven asymptotic convergence of the spectral norm. At best, we may hope that the consistency in the spectral norm proven in [20] suggests consistency of the dominant "spiky" eigenvalues estimation with $\boldsymbol{\lambda}_j > 1$.

Since the analytical conditions for production of negative eigenvalues by redundancy averaging remain unspecified by RMT with the worst condition (17) that follows from the existing results, we introduce below the results of Monte-Carlo simulations. These simulation results should illustrate the negative eigenvalues production capabilities of the redundancy averaging, as well as the properties of the "spiked" and "non-spiked" eigenvalues for different $N, T, T/N = c$.

For the matrix (8)-(10), $N = 17$ and $T = 85$ $(T/N = 5)$, the eigenvalues of the sample covariance matrix $\widehat{\mathbf{R}}_N$ and the eigenvalues of the redundancy averaged matrix $\widehat{\mathbf{T}}_N^{\mathrm{ra}}$ are also introduced in TABLE II.

TABLE II.

| $\lambda(\boldsymbol{T}_N)$ | $\lambda(\widehat{\boldsymbol{R}})$ | $\lambda(\widehat{\boldsymbol{T}}_N^{ra})$ |
|---|---|---|
| 1.496411 | 1.992888 | 2.314267 |
| 1.424830 | 1.533716 | 1.348330 |
| 1.136760 | 1.336132 | 1.151430 |
| 1.000872 | 1.121411 | 0.943905 |
| 1.000083 | 0.999965 | 0.788519 |

| $\lambda(\boldsymbol{T}_N)$ | $\lambda(\widehat{\boldsymbol{R}})$ | $\lambda(\widehat{\boldsymbol{T}}_N^{ra})$ |
|---|---|---|
| 0.992370 | 0.833157 | 0.612460 |
| 0.814989 | 0.709042 | 0.518154 |
| 0.450598 | 0.314187 | 0.480149 |
| 0.158410 | 0.139769 | 0.399148 |
| 0.023622 | 0.015833 | 0.382318 |
| 0.002041 | 0.001712 | 0.295009 |
| 0.000210 | 0.000233 | 0.225881 |
| 0.000104 | 0.000102 | 0.074540 |
| 0.000100 | 0.000086 | -0.062338 |
| 0.000100 | 0.000066 | -0.086109 |
| 0.000100 | 0.000059 | -0.142516 |
| 0.000100 | 0.000051 | -0.244740 |

In Fig. 1 we introduce the sample probability density function (pdf) of the 4 smallest eigenvalues $\widehat{\boldsymbol{\lambda}}_{14}$ to $\widehat{\boldsymbol{\lambda}}_{17}$ averaged over 1000 trials.

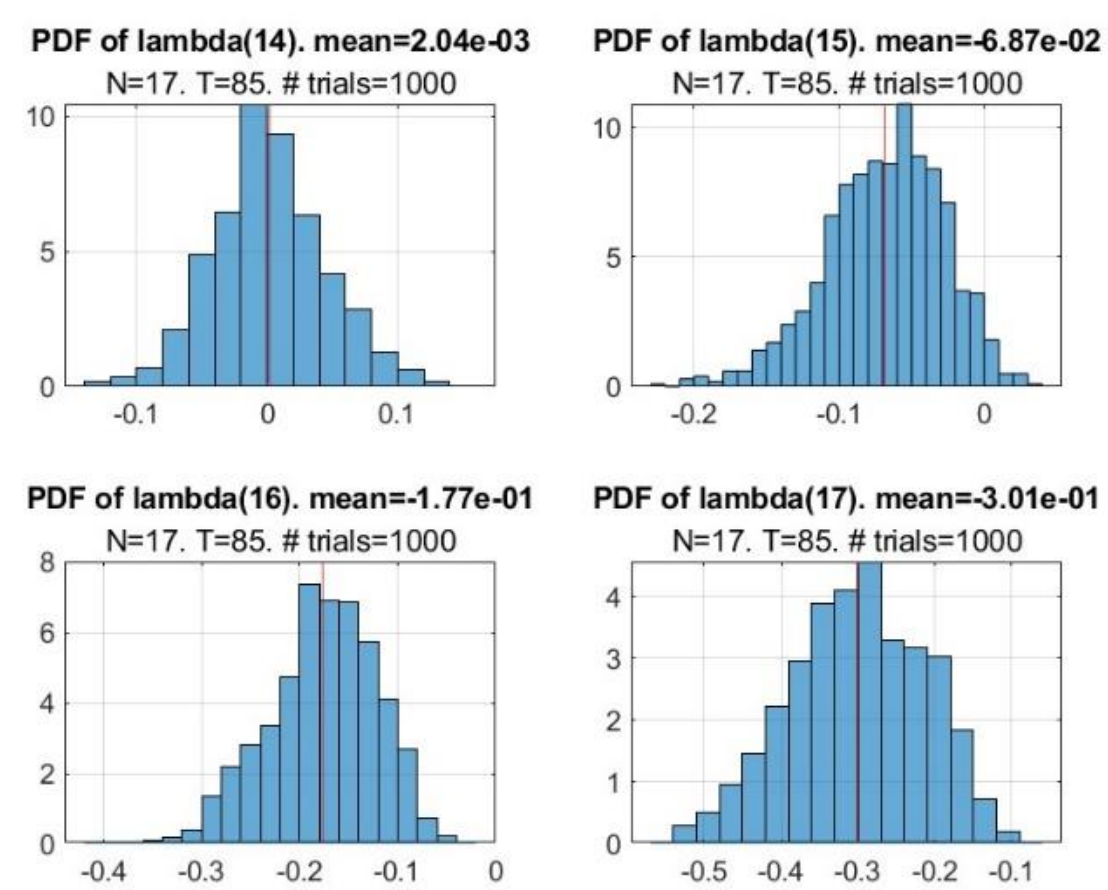

Fig. 1. PDF of the last 4 eigenvalues of the sample matrix. N=17, T=85.

One can see that the last five eigenvalues of the exact Toeplitz Hermitian matrix are ranging from $1.04 \cdot 10^{-4}$ to $1.00 \cdot 10^{-4}$. If we use (17) for the minimum support $T$ evaluation, required for the last eigenvalue $\boldsymbol{\lambda}_{17}$ to stay positive, then this sample volume should exceed $10^8$ (!!!) samples. Not surprisingly, the sample pdf of the minimum eigenvalue $\widehat{\boldsymbol{\lambda}}_{17}$ of the redundancy averaged matrix for $T = 85$ fully resides in the negative domain.

Now let us analyze the dynamics of the negative eigenvalues when $N \to \infty$, $T \to \infty$, and $T/N \to c$. For this analysis we considered the cases with $N = 51, T = 255$ $(T/N = 5)$, and $N = 102, T = 510$ $(T/N = 5)$. The results of these simulations are presented in Fig. 2 and Fig. 3.

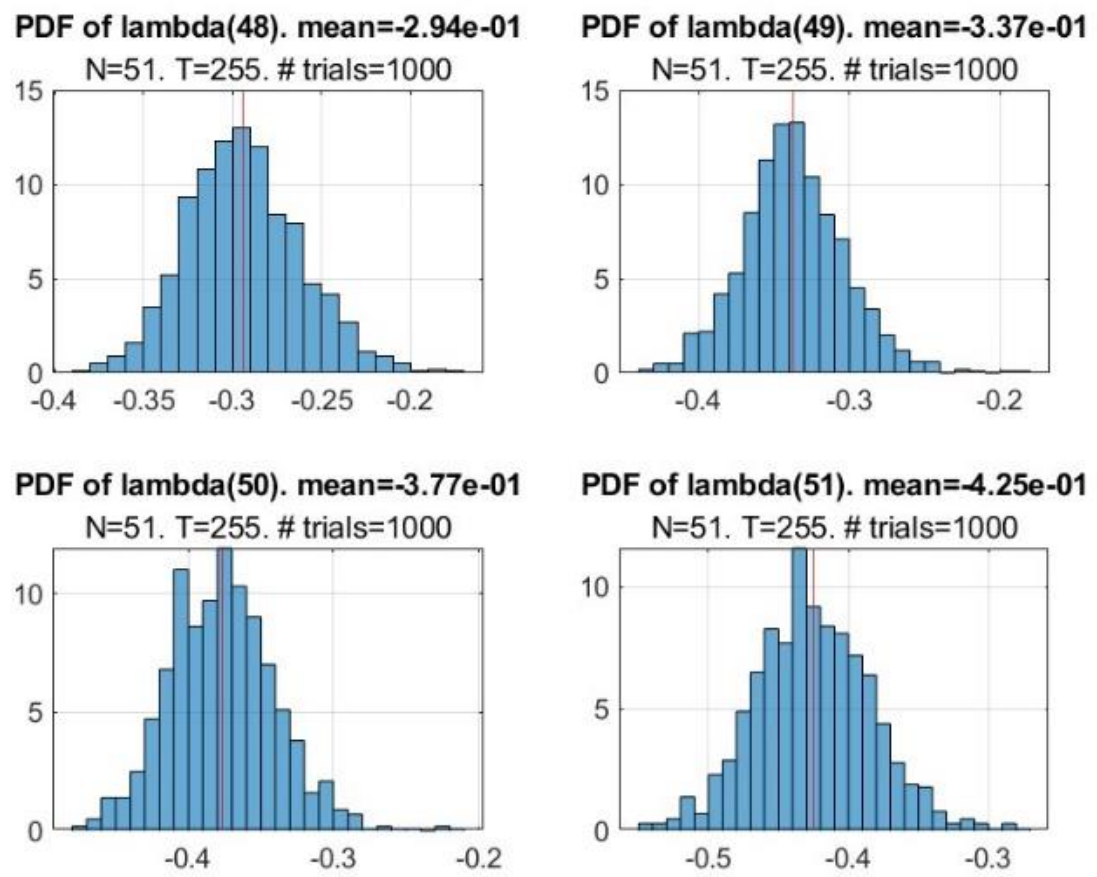

Fig. 2. PDF of the last 4 eigenvalues of the sample matrix. N=51, T=255.

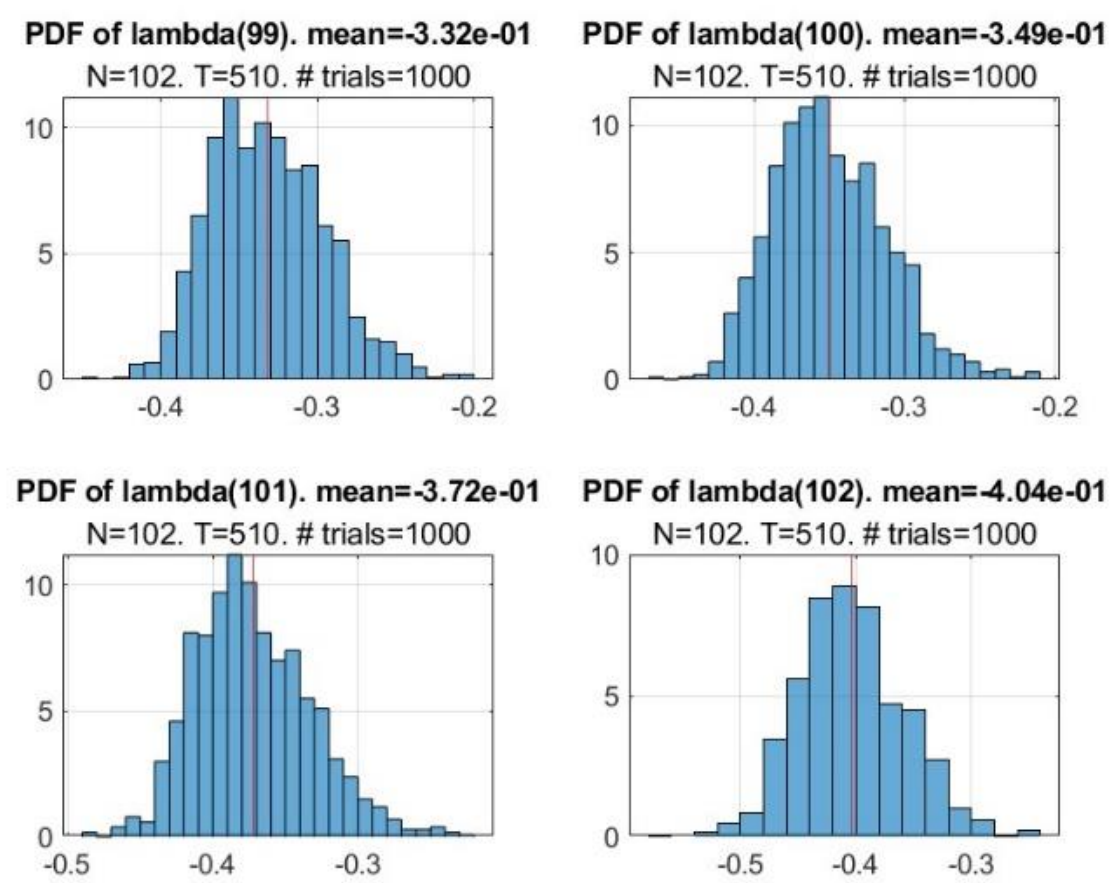

Fig. 3. PDF of the last 4 eigenvalues of the sample matrix. N=102, T=510.

One can see, that with $N$ moving from $N = 17$ to $N = 102$ ($T = 85$ to $T = 510$), the minimum eigenvalue remains negative, even shifted a bit deeper into the negative domain. Let us analyze other property dynamics on the root from $N = 17$ to $N = 102$, while keeping $c = T/N = 5$ the same. In Fig. 4 we present the pdf's, averaged over 1000 trials of the spectral norm of the redundancy averaged Toeplitz matrix ($\|\mathbf{T}_N - \mathbf{T}_N^{\mathrm{ra}}\|$ for the case $N = 17, T = 85; N = 51, T = 255;$ and $N = 102, T = 510$ correspondingly).

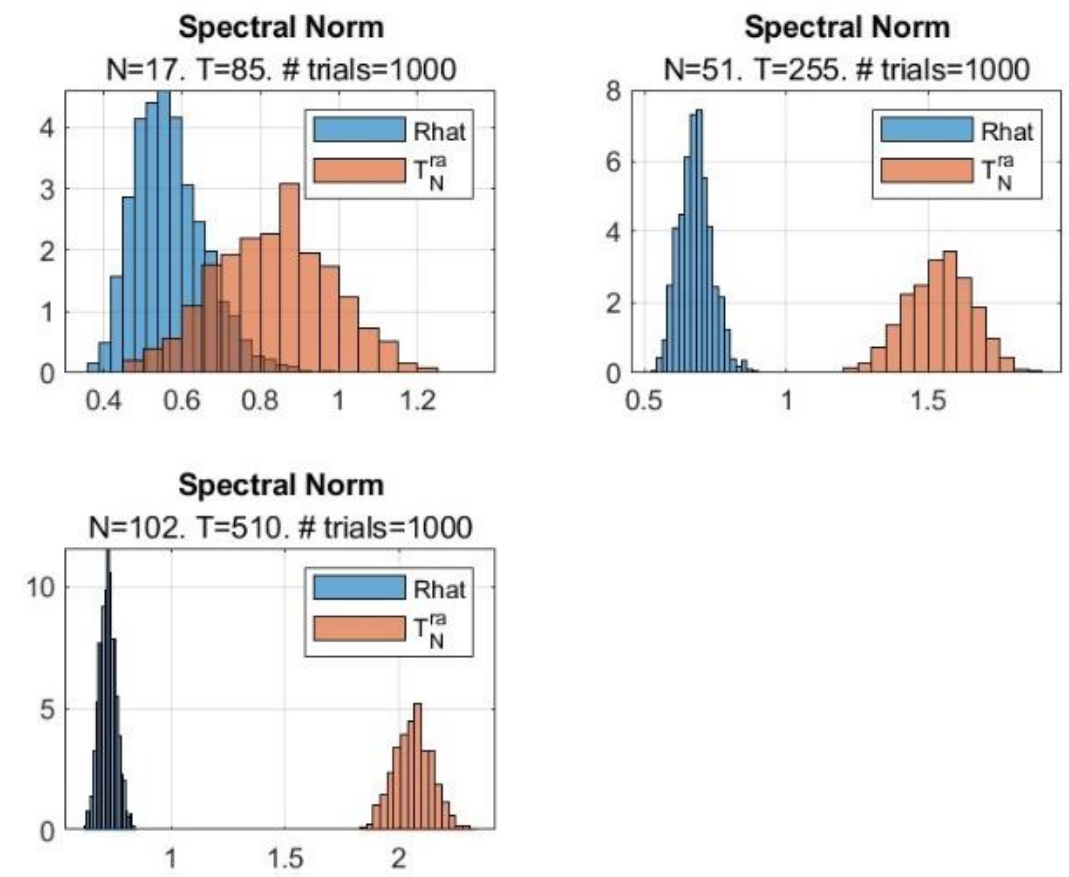

Fig. 4. Spectral norm of the sample matrix. T/N = 5.

Once again, the spectral norm is found to be growing in transition from $N = 17, T = 85$ to $N = 102, T = 510$. The standard sphericity test that does not exist (does not make sense) for negative eigenvalues, ranges from (0.1 - 0.25) for $N = 17, T = 85$, and within the range $(0.25 – 6.2)\cdot 10^{-5}$ for $N = 51, T = 255$. Finally, for $N = 102, T = 510$ this likelihood ratio is distributed within the range $(1 - 3)\cdot 10^{-5}$ (!!!) over this transition. Note, that the spectral norm does not converge to zero. It is important to know that the minimalist diagonal loading of the redundancy averaged Toeplitz covariance matrix does not improve the likelihood of the loaded solution. In our example, the loading that resulted in the minimal eigenvalue, equal to the RMT-advised minimum eigenvalue, had the sphericity likelihood ratio equal to $2.3\cdot 10^{-25}$.

Finally, let us introduce the "spiked" sphericity test:

$$\mathrm{LR}(\widehat{\boldsymbol{u}}_{N-m}, \widehat{\mathbf{R}}_N) = \frac{\prod_{j=m+1}^{N} \widehat{\boldsymbol{u}}_j^{\mathrm{H}} \widehat{\mathbf{R}}_N \widehat{\boldsymbol{u}}_j}{[\frac{1}{N-m} \sum_{j=m+1}^{N} \widehat{\boldsymbol{u}}_j^{\mathrm{H}} \widehat{\mathbf{R}}_N \widehat{\boldsymbol{u}}_j]^{N-m}} \tag{20}$$

where $\widehat{\boldsymbol{u}}_j$ are the eigenvectors of the redundancy averaged matrix, corresponding to the “non-spiky” eigenvalues. For all "non-spiky" eigenvalues that correspond to the matrix's noise subspace, which is the last four (or five) eigenvalues of the matrix $\mathbf{T}_N$ (8)-(10), for $\boldsymbol{u}_j, j = m+1, \ldots, N$ of the true covariance matrix, when

$$\boldsymbol{u}_j^{\mathrm{H}} \widehat{\mathbf{R}}_N \boldsymbol{u}_j = \boldsymbol{\lambda}_j \boldsymbol{u}_j^{\mathrm{H}} \widehat{\mathbf{S}}_T \boldsymbol{u}_j, \tag{21}$$

the likelihood ratio does not depend on $\mathbf{T}_N$ since the pdf of $\gamma_o$

$$\gamma_o = \mathrm{LR}(\widehat{\boldsymbol{u}}_{N-m}, \widehat{\mathbf{R}}_N) = \frac{\prod_{j=m+1}^{N} \boldsymbol{u}_j^{\mathrm{H}} \widehat{\mathbf{S}}_T \boldsymbol{u}_j}{[\frac{1}{N-m} \sum_{j=m+1}^{N} \boldsymbol{u}_j^{\mathrm{H}} \widehat{\mathbf{S}}_T \boldsymbol{u}_j]^{N-m}} \tag{22}$$

$$\mathbf{S}_T = \frac{1}{T} \sum_{t=1}^{T} \boldsymbol{\xi}_t \boldsymbol{\xi}_t^{\mathrm{H}}, \quad \boldsymbol{\xi}_t \sim \mathbb{C}\mathcal{N}(0, \mathbf{I}_N)$$

is specified by $(N, T)$ only.

Indeed, the independence is due to orthogonality of the eigenvectors $\boldsymbol{\mathcal{U}}_j$ and the random numbers $\boldsymbol{\mathcal{U}}_j^{\mathrm{H}}\hat{\mathbf{S}}_T\boldsymbol{\mathcal{U}}_j$ are the values from the Marchenko-Pastur distribution [26]:

$$f_\beta(x) = \frac{\sqrt{(x-\lambda_{min})^+(\lambda_{max}-x)^+}}{2\pi\beta x} \tag{23}$$

where

$$\frac{N}{T} = \beta, \quad (z)^+ = \max(0, z), \tag{24}$$

$$\lambda_{min} = (1-\sqrt{\beta})^2, \ \ \lambda_{max} = (1+\sqrt{\beta})^2 \tag{25}$$

Obviously, for the true maximum likelihood estimate $\hat{\mathbf{T}}_N^{\mathrm{ML}}$,

$$\mathrm{LR}\big(\boldsymbol{\mathcal{U}}_{N-m}(\hat{\mathbf{T}}_N^{\mathrm{ML}}), \hat{\mathbf{R}}_N\big) > \gamma_o \tag{26}$$

and for the small $(N-m) \ll T$, the pdf for $\mathrm{LR}\big(\hat{\boldsymbol{\mathcal{U}}}_{N-m}, \hat{\mathbf{R}}_N\big)$ should be very close to one.

Yet, for the redundancy averaged covariance matrix $\mathbf{T}_N^{\mathrm{ra}}$ for $N = 17, T = 85$ (Fig. 5), $N = 51, T = 255$ (Fig. 6), and $N = 102, T = 510$ (Fig. 7), this "spiky" LR remains almost equally distributed within the range $0 \le \mathrm{LR} \le 0.7$.

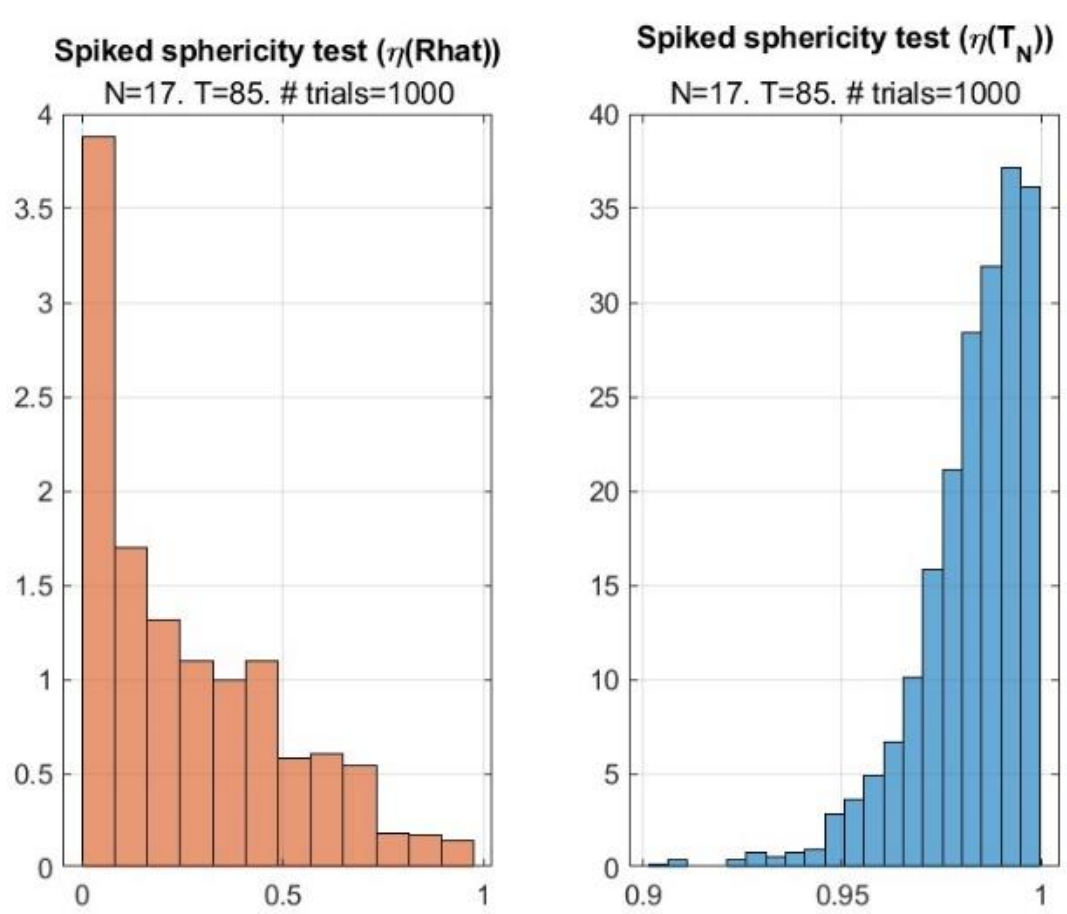

Fig. 5.Spiked sphericity test. Sample matrix (left) and true matrix (right). N=17, T=85.

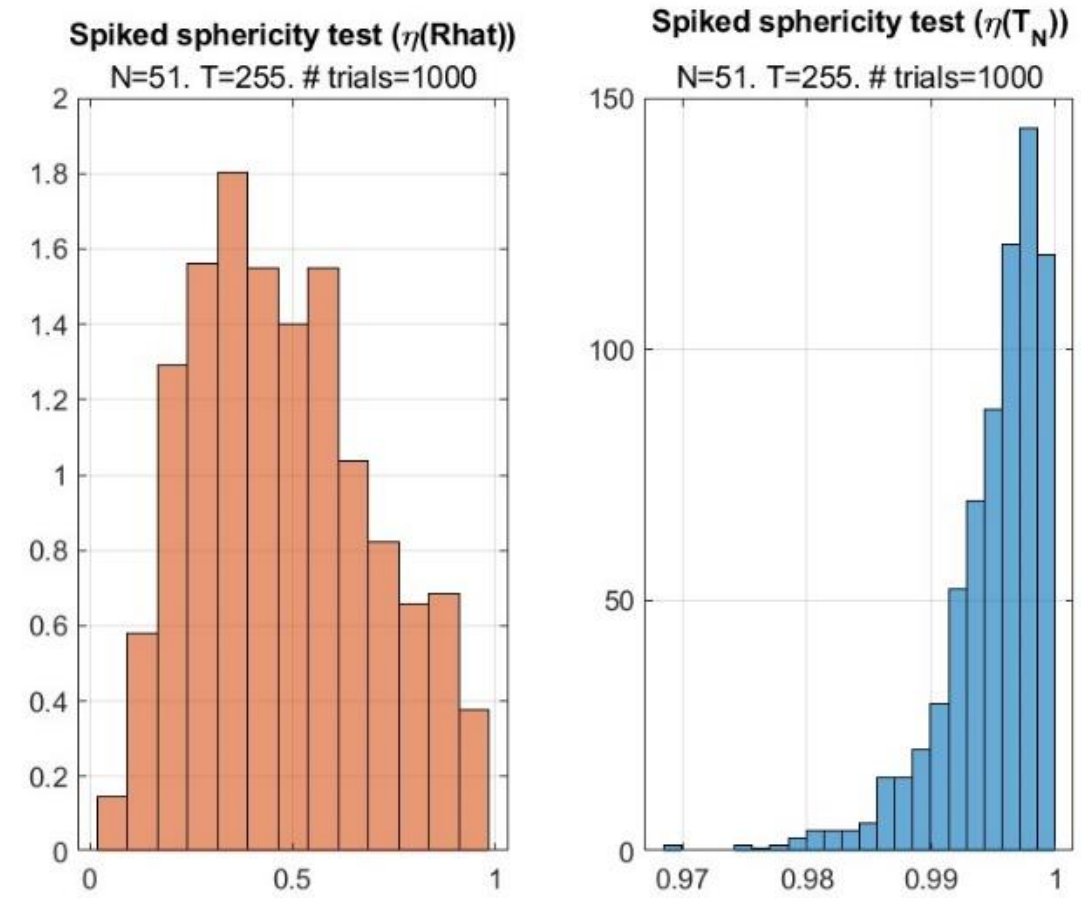

Fig. 6.Spiked sphericity test. Sample matrix (left) and true matrix (right). N=51, T=255.

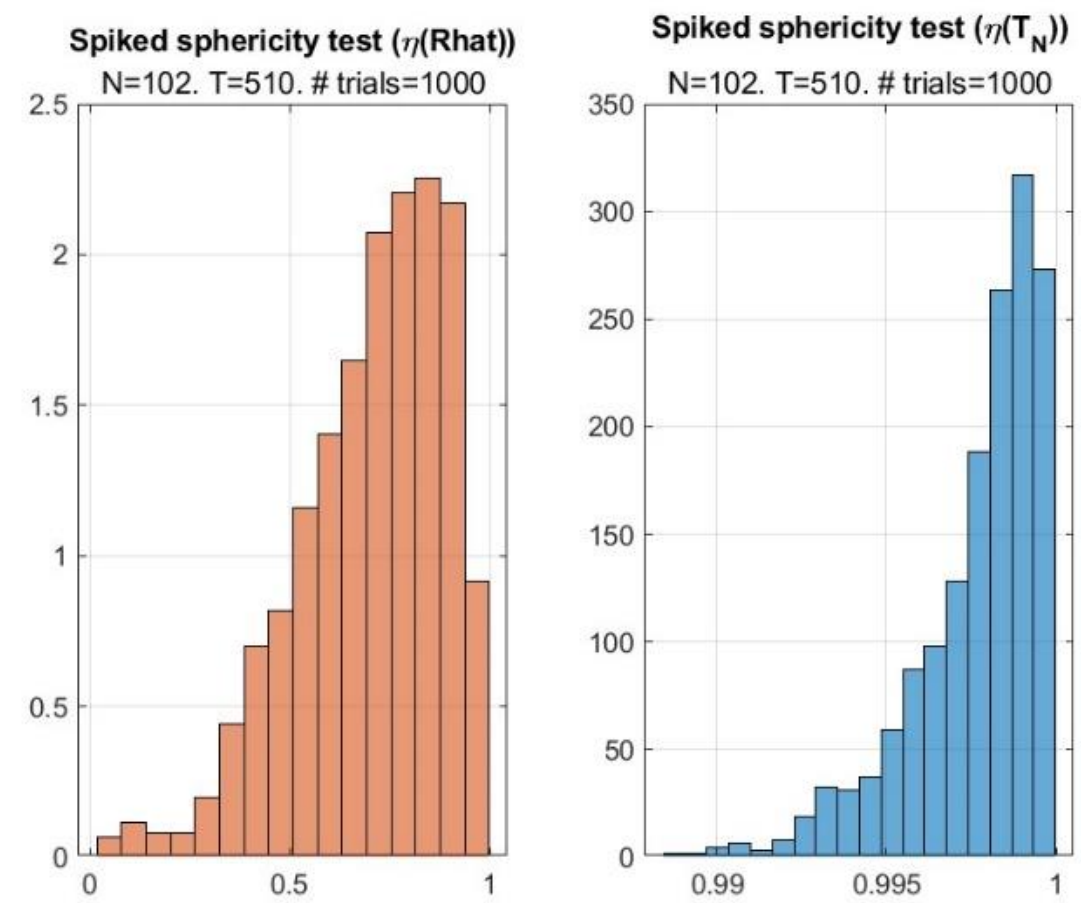

Fig. 7.Spiked sphericity test. Sample matrix (left) and true matrix (right). N=102, T=510.

Therefore, at least for the tested $(N, T)$ values with the same $T/N = 5$ ratio, all tested criteria remain the same or even smaller than for the original case with $N = 17, T = 85$. It is very important that the "full size" sphericity likelihood ratio of the redundancy averaged and rectified covariance (i.e. with no negative eigenvalues) remains several orders of magnitude (!) smaller than the LR values generated by the true covariance matrix. Therefore, for the fixed ratio $T/N = 5$, the redundancy averaging provides solutions very far from the maximum-likelihood estimation (MLE), at least for the analyzed typical scenario. Naturally, the next question is on the traditional convergence with $N = \mathrm{const}, T \to \infty$ $(T/N \to \infty)$. This analysis was conducted for all three antenna dimensions $N = 17, 51, 102$ for $T/N \ge 10$.

Analysis of the negative last four eigenvalues for $N = 17, T = 170$ (Fig. 8) demonstrated no significant difference in their distributions compared with the case $N = 17, T = 85$.

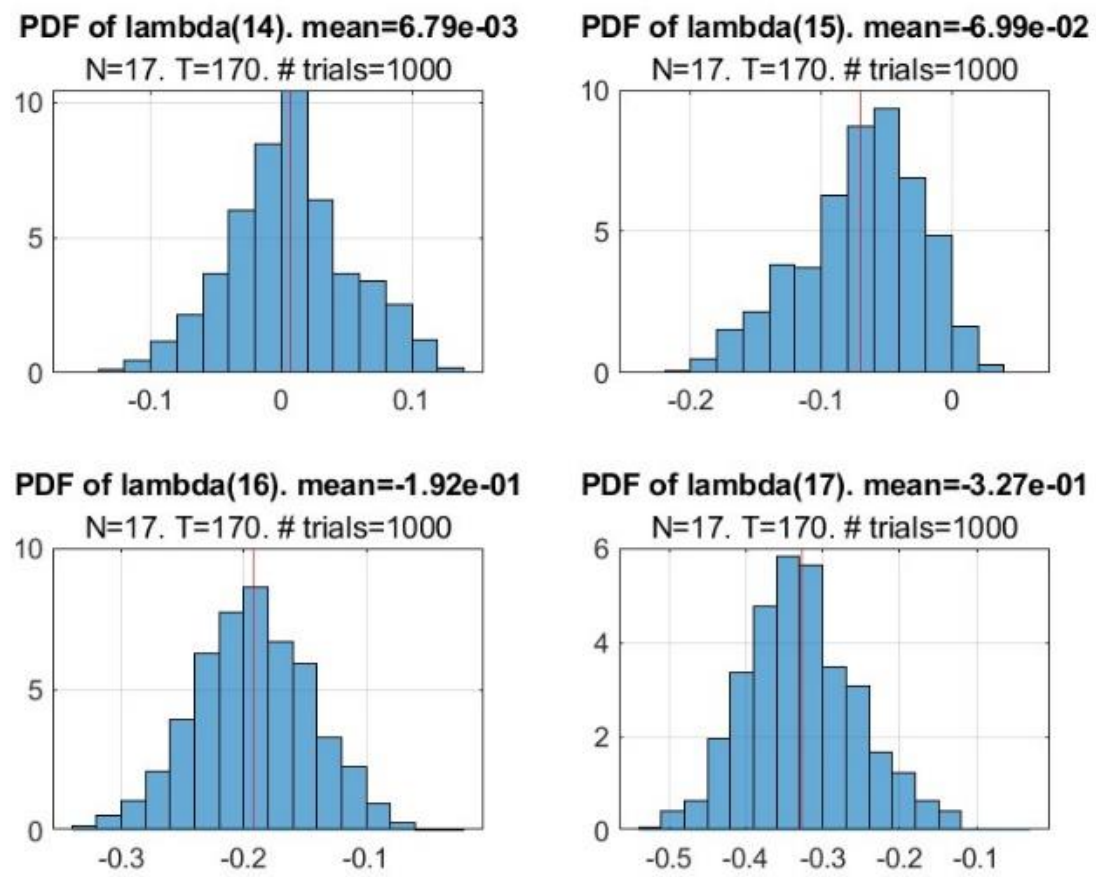

Fig. 8. PDF of the last 4 eigenvalues of the sample matrix. N=17, T=170.

The same insignificant changes are observed for $N = 51, T = 510$ and $N = 102, T = 1020$. The spectral norm $\|\mathbf{T}_N - \mathbf{T}_N^{\text{ra}}\|$ (Fig. 9) improves proportionally as the sample volume is doubled.

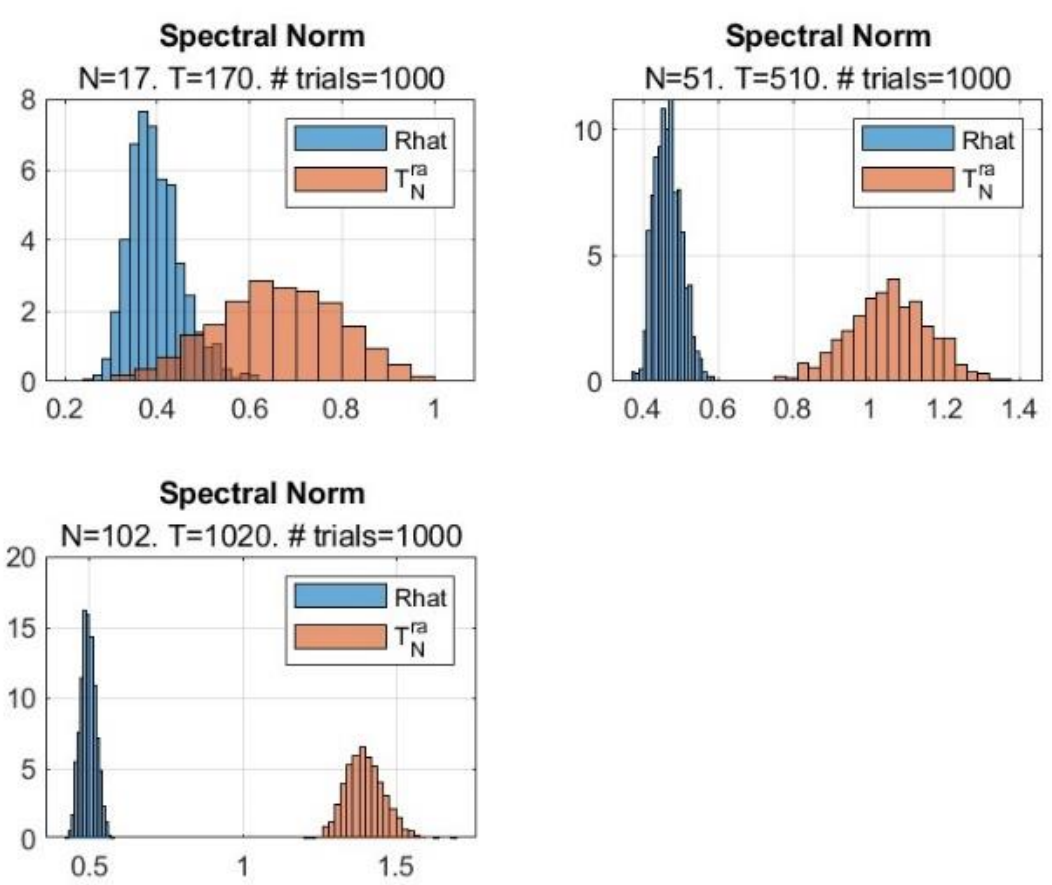

Fig. 9. Spectral norm of the sample matrix. T/N = 10.

The regular sphericity test became several orders of magnitude bigger, but still remained many orders of magnitude below the LR values generated by the true covariance matrix $\mathbf{T}_N$ (8)-(10). The "spiked" sphericity test (Fig. 10-Fig. 12) remained practically unchanged compared with the original case ($N = 17, T = 85$).

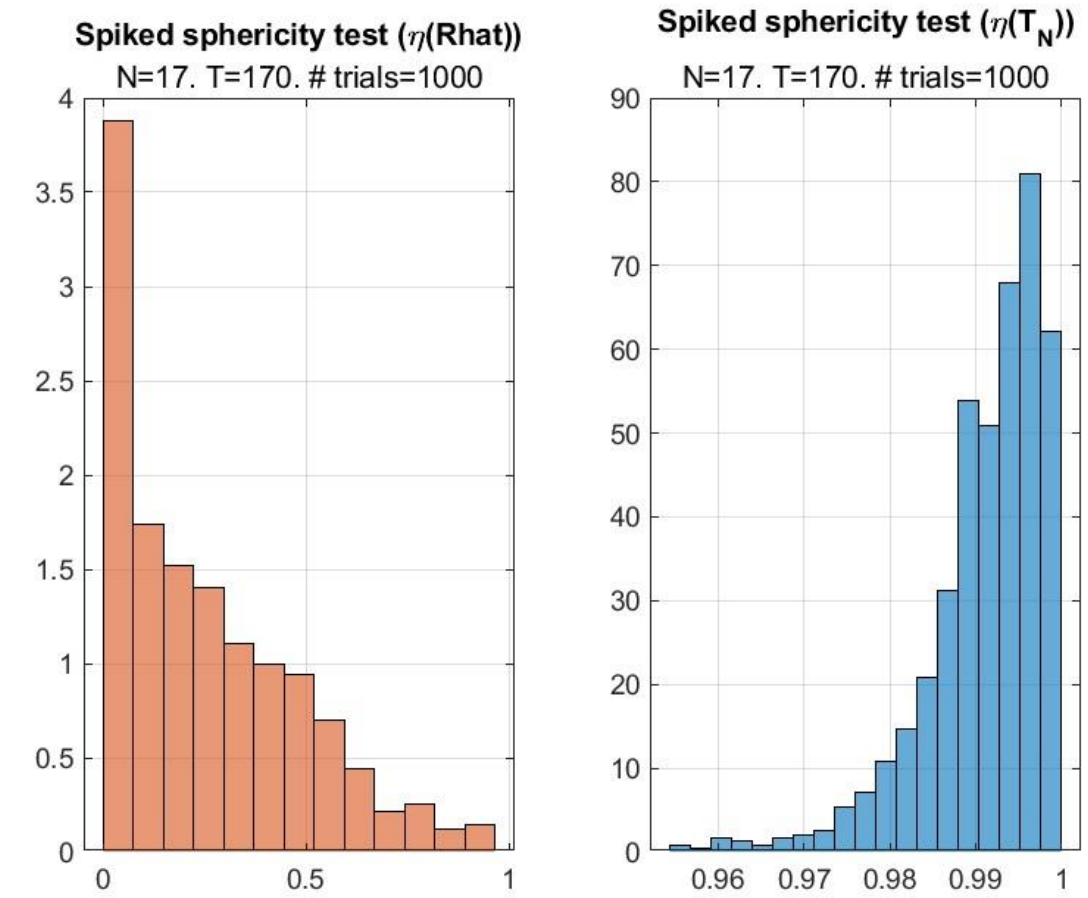

Fig. 10.Spiked sphericity test. Sample matrix (left) and true matrix (right). N=17, T=170.

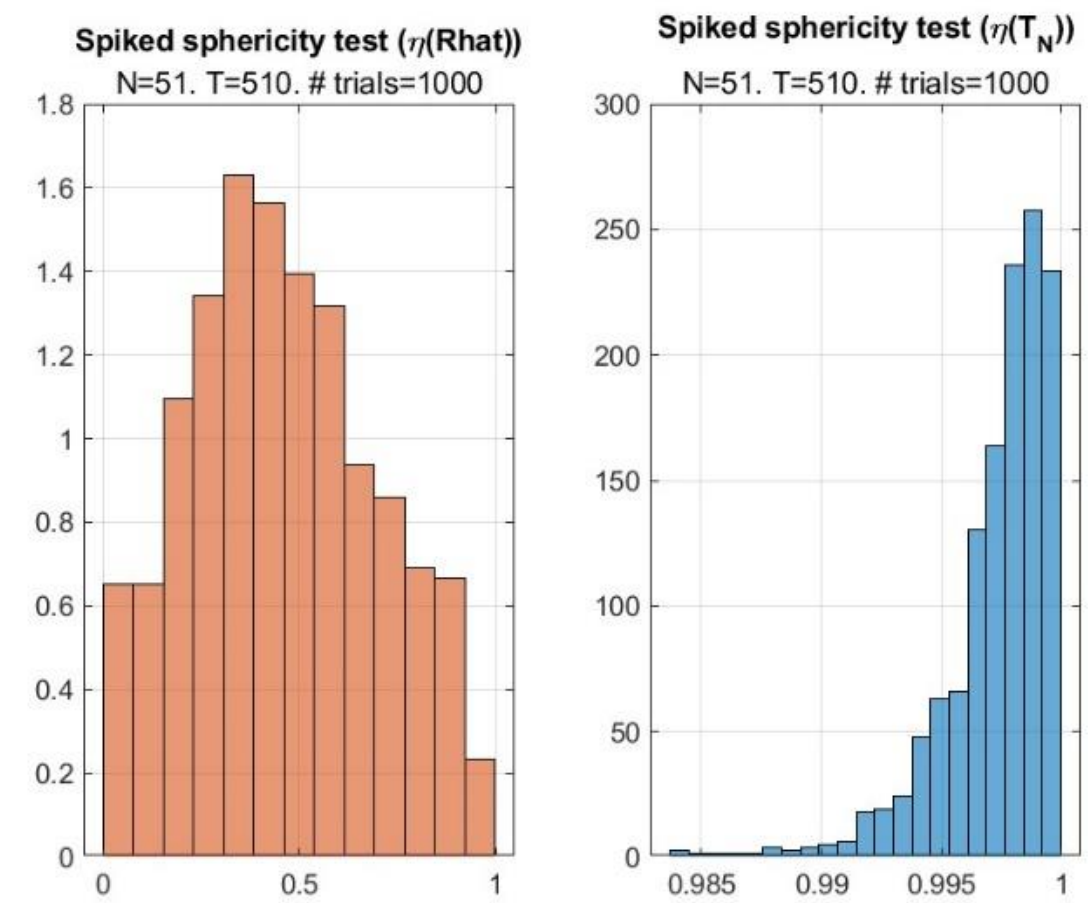

Fig. 11.Spiked sphericity test. Sample matrix (left) and true matrix (right). N=51, T=510.

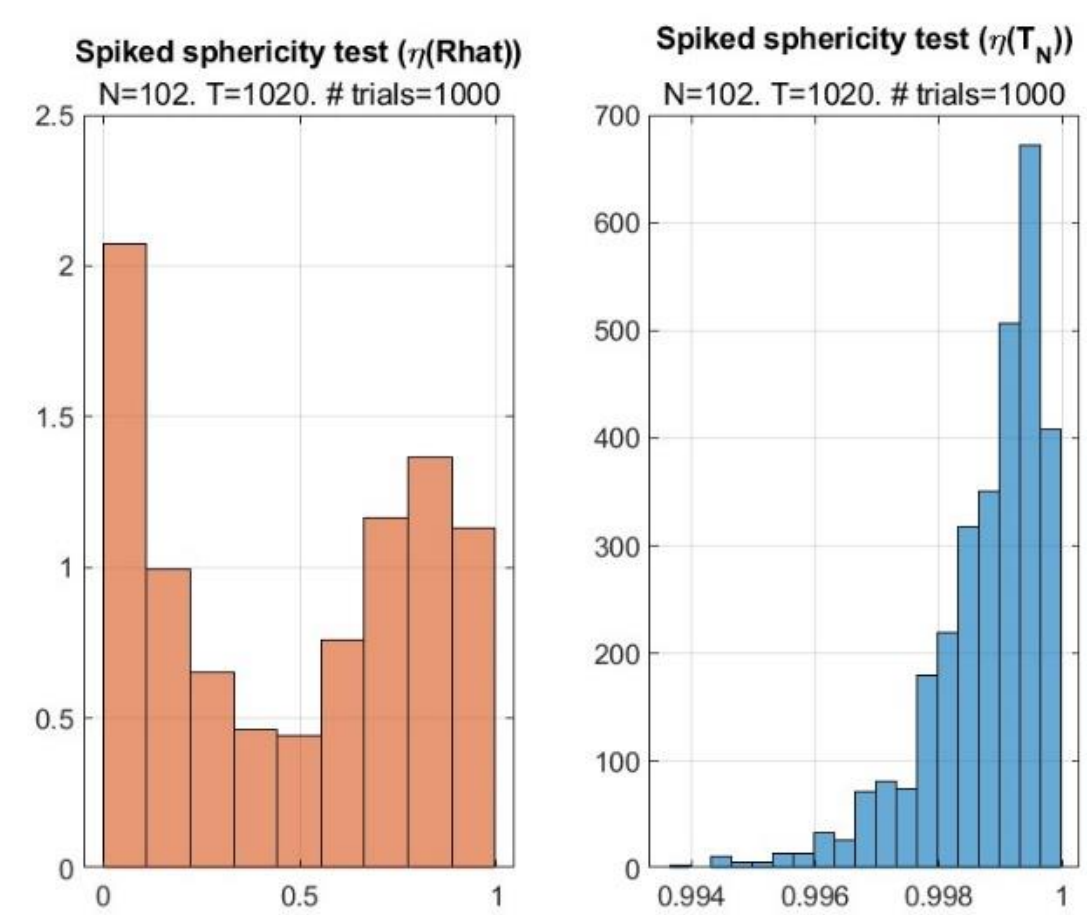

Fig. 12.Spiked sphericity test. Sample matrix (left) and true matrix (right). N=120, T=1020.

Skipping all interim results, let us report on the case $N = 17, T = 10^7$ (!). The sample distributions of the last four eigenvalues, introduced in Fig. 13 demonstrates that only the fourth smallest eigenvalue $\hat{\boldsymbol{\lambda}}_{14}$ became positive, while the last (17th) eigenvalue remained mostly negative.

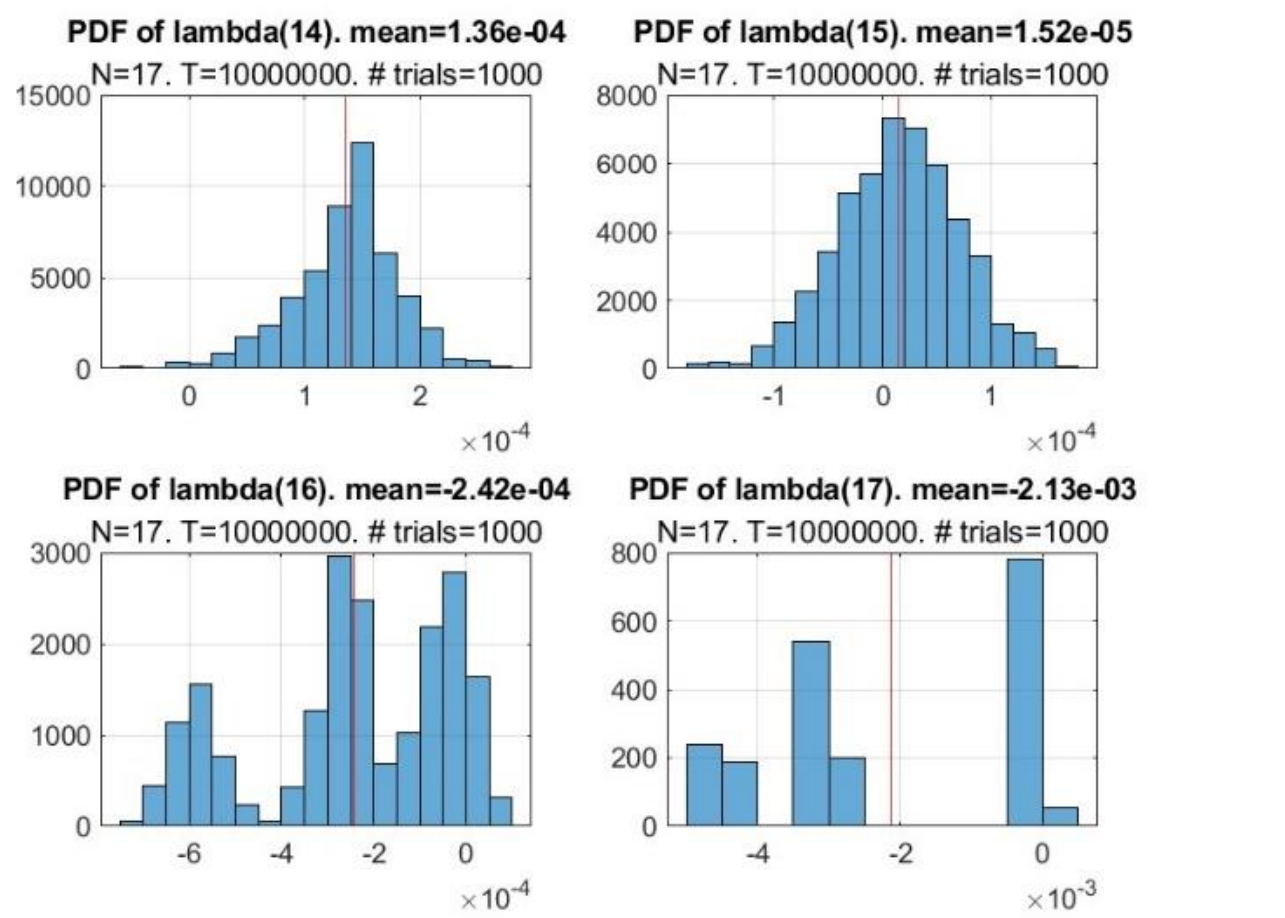

Fig. 13. PDF of the last 4 eigenvalues of the sample matrix. N=17, T=$10^7$.

While the spectral norm improved, reaching $10^{-3}$ (Fig. 14), both the regular and "spiky" sphericity tests remained outside of the pdf of the LR values, calculated for the true covariance matrix. (Fig. 15 and Fig. 16) even for this enormously large sample volume.

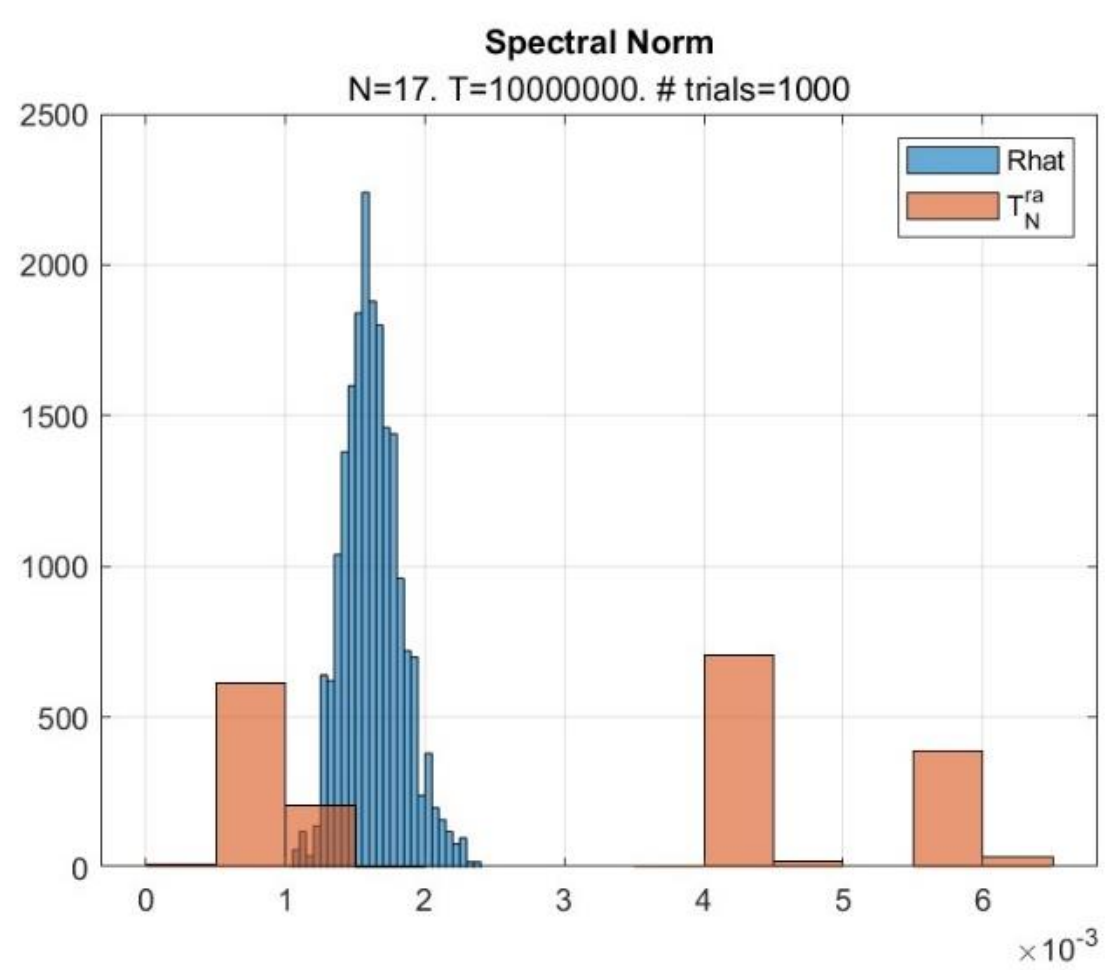

Fig. 14. Spectral norm of the sample matrix. N = 10, T = $10^7$.

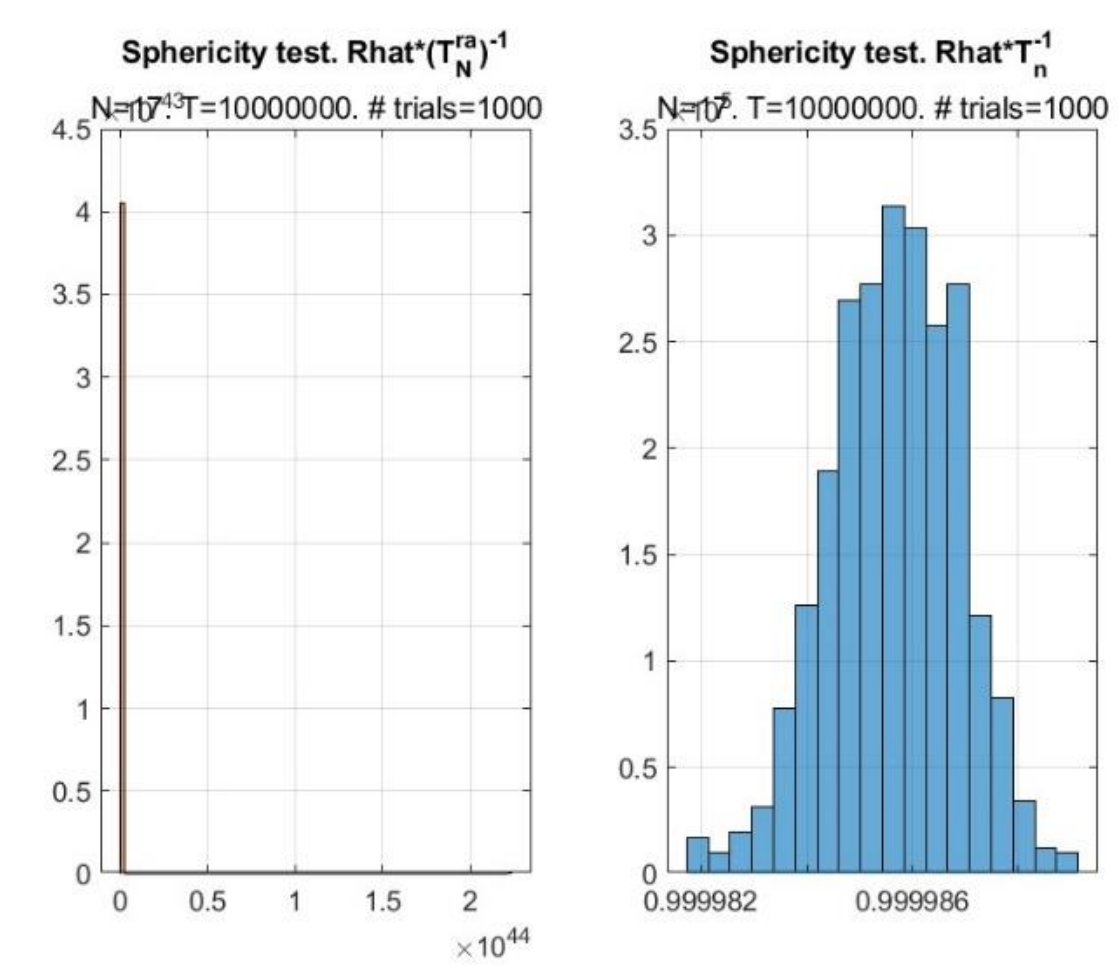

Fig. 15.Normal sphericity test. Sample matrix with redundancy averaged sample matrix (left) and true matrix (right). N=17, T=$10^7$.

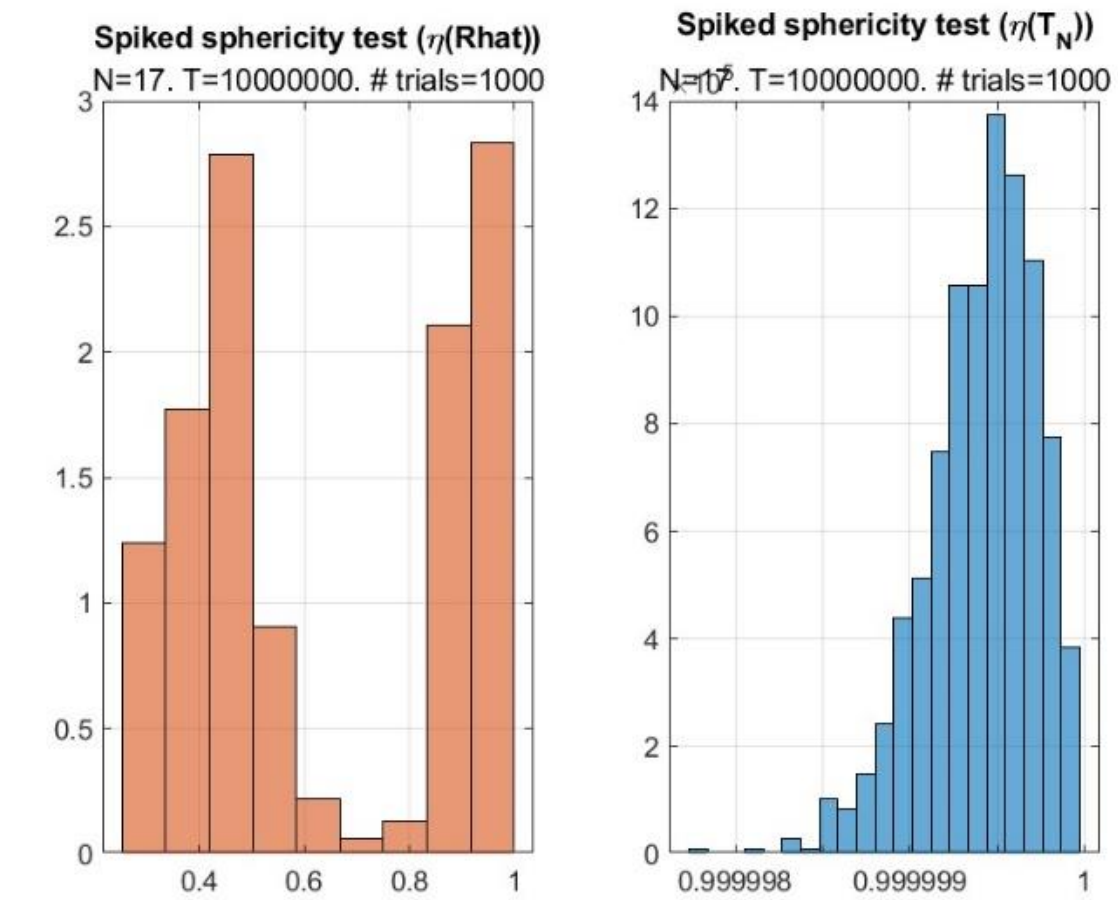

Fig. 16.Spiked sphericity test. Sample matrix (left) and true matrix (right). N=17, T=$10^7$.

This detailed analysis of the redundancy averaging routine applied for the "spiked" covariance matrix with the realistic (~30 dB) input signal-to-noise ratio demonstrated that the convergence of the redundancy averaged matrix to the true Toeplitz Hermitian covariance matrix proven in [6], [20] should be treated with caution, since for the "spiked" covariance matrices and the parameters $N$, $T$ changing within the significant range, when $N \to \infty$, $T \to \infty$, $T/N = c$, this convergence may not be observed for the important properties of the matrix.

Finally, these simulations confirm that the condition for the sample volume required to avoid negative eigenvalues by redundancy averaging is indeed close to (17). To check this condition, we increased the noise power $\sigma_n^2 = 10^{-2}$ (from $10^{-4}$) so that the minimum sample volume $T_{min}$

$$T_{min} \forall \frac{N}{T\lambda_{min}^2} \ll 1, \quad T_{min} = 2 \cdot 10^5 \qquad (27)$$

The sample pdf over 1000 trials for the $\hat{\lambda}_{min}$ of the redundancy averaged Toeplitz matrix and supporting statistics are illustrated in Fig. 17-Fig. 20. This data demonstrates a zero (!) probability of the negative eigenvalue generation.

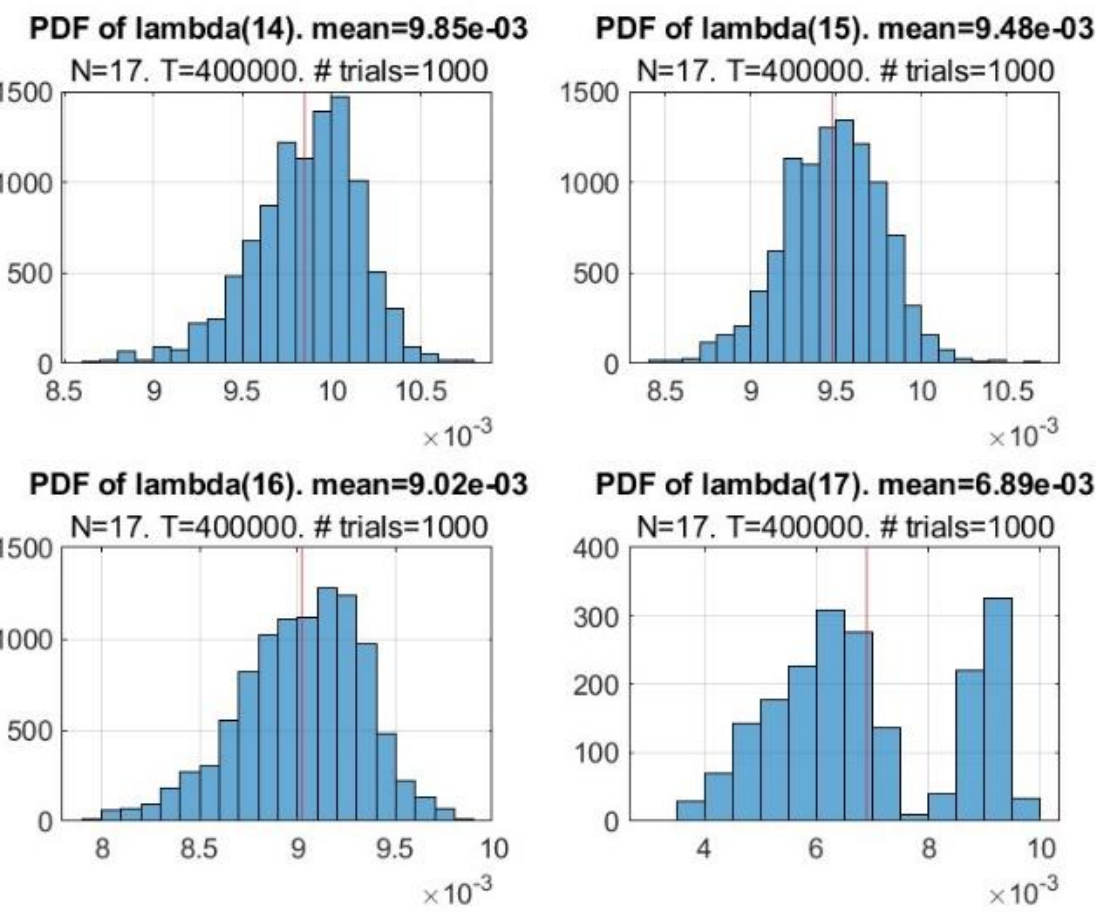

Fig. 17. PDF of the last 4 eigenvalues of the sample matrix. N=17, T=4·10[5], $\sigma_n^2 = 10^{-2}$.

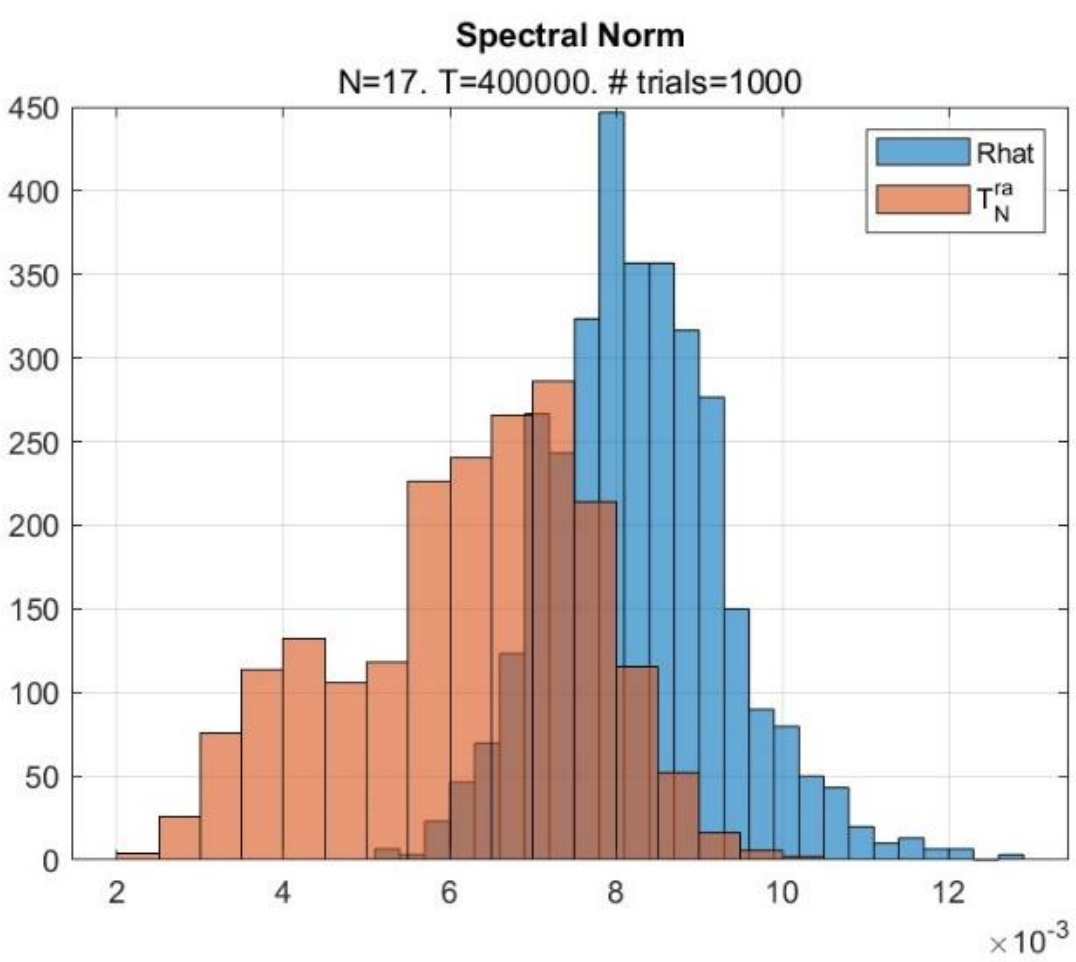

Fig. 18. Spectral norm of the sample matrix. N = 17, T = 4·10[5]. $\sigma_n^2 = 10^{-2}$.

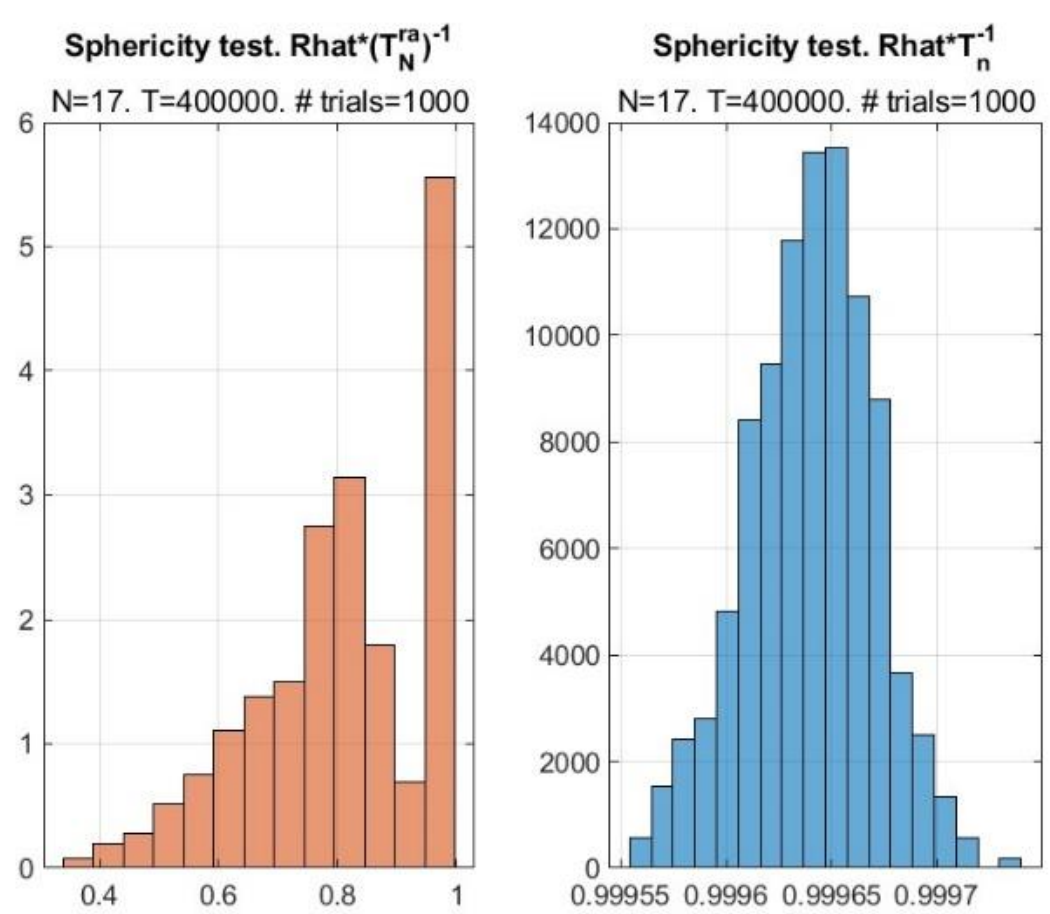

Fig. 19. Normal sphericity test. Sample matrix with redundancy averaged sample matrix (left) and true matrix (right). N=17, T=4·10[5]. $\sigma_n^2 = 10^{-2}$.

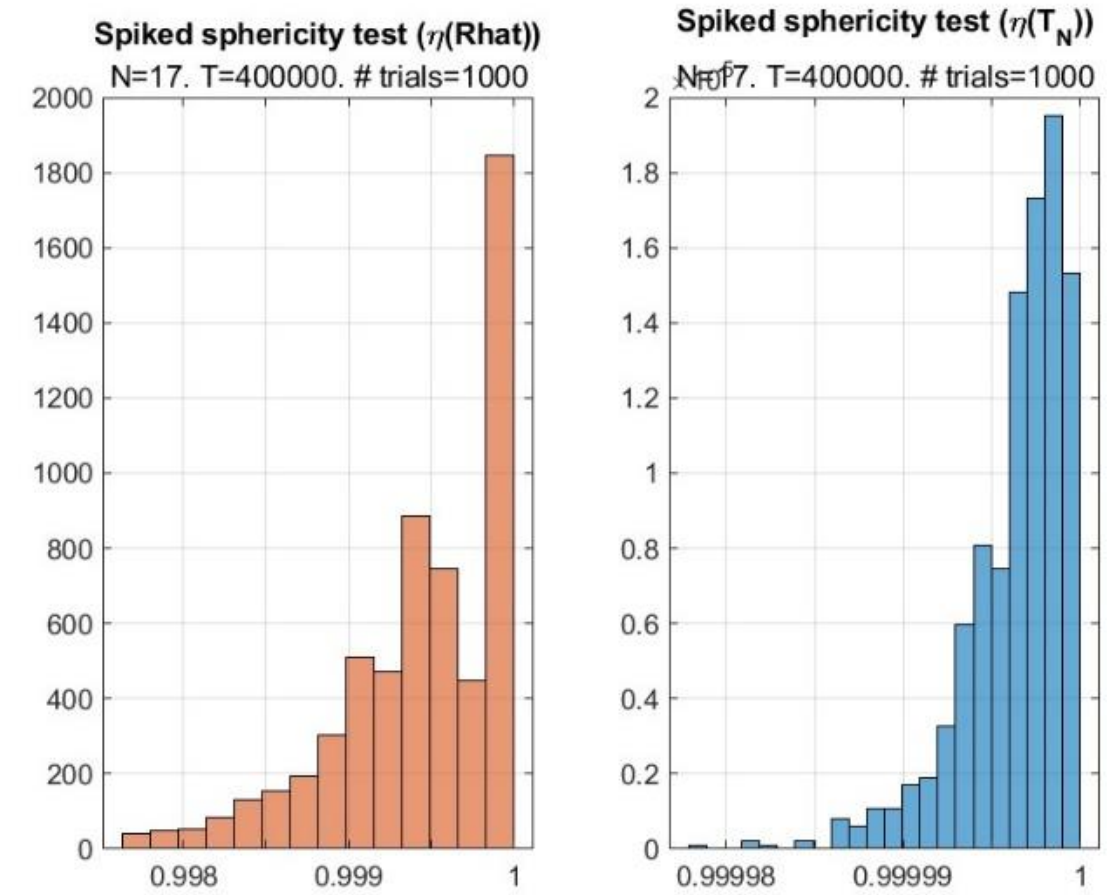

Fig. 20. Spiked sphericity test. Sample matrix (left) and true matrix (right). N=17, T=4·10[5]. $\sigma_n^2 = 10^{-2}$.

## III. Rectification and Alternative to the Redundancy Averaging for Production of the ML-Proxy Hermitian Toeplitz Covariance Matrix Estimate

For generation of the ML-proxy p.d. Toeplitz Hermitian covariance matrix estimate we may either consider rectification of the non-p.d. Toeplitz matrix, produced by redundancy averaging, or consider an alternative technique for transition from the p.d. Hermitian (non-Toeplitz) sample matrix to a p.d. Hermitian Toeplitz matrix for any reasonable sample volume. In this section we introduce these techniques, while in the next section we provide the results of the Monte-Carlo simulation.

Let us start from the rectification routine. The main problem here is the uniqueness of the Hermitian matrix reconstruction. Indeed, given a p.d. sample Hermitian matrix, it is quite logical to try to rectify the non-p.d. redundancy averaged Toeplitz matrix into a p.d. matrix with eigenvalues that are the same or modified by RMT. But this problem does not have a unique solution. In [23] it was proven that even within the class of real-valued Toeplitz matrices this solution is not unique. In [22] an example of the Toeplitz Hermitian matrix generation with the given eigenvalues was introduced with no initial solution provided. Therefore, we had to request a Hermitian Toeplitz matrix with the given eigenvalues having a minimal distance (in whatever norm) from the redundancy averaged. Another option would be to fix the norm where the optimized Toeplitz matrix differs from the redundancy averaged, and minimize the difference in the all positive eigenvalues of the optimized matrix and the prescribed set of eigenvalues. Unfortunately, we are not aware of such optimization techniques and have to rely upon the expectation that by improving the eigenvalues of the redundancy averaged matrix, we may find a solution with a minimal norm of its difference with the redundancy averaged ones. Therefore in pursuing this approach, we first had to specify the positive eigenvalues of the reconstructed Toeplitz matrix, and then propose an iterative technique to move from the initial to the optimum solution.

In specifying the eigenvalues of the reconstructed Toeplitz matrix, we first have to estimate its "noise-subspace" dimension (non-spiked eigenvalues) in the sample matrix $\widehat{\mathbf{R}}_N$. This is one of the classical signal processing problems, addressed by the so-called "Information Theoretic Criteria", such as MDL or AIC [8]. We apply our "Expected Likelihood" approach [9] when the number of noise subspace eigenvalues is found from the condition that the sample matrix with the modified eigenvalues has statistically the same likelihood as the true (unknown) covariance matrix. For the specified number of "noise subspace" (non-spiked) eigenvalues, we may apply the RMT methodology [11] for the eigenvalues rectification.

Given this set of eigenvalues we want to achieve, we now have to modify the moduli and phases of the covariance lags of the redundancy averaged matrix. Note that the existing mathematical Toeplitz Inverse Eigenvalue Problem (ToIEP) methodology considers only the real-valued problem of this kind due to the identifiability problems discussed above. For this reason, we had to modify the existing routines [14] with the alternating ($N$ - 1) phases and $N$ moduli optimization, with each stage of this optimization having the right number ($\leq N-1$) of modified parameters.

While we could start our optimization from the redundancy averaged matrix with its negative eigenvalues, we first converted the non-p.d. redundancy averaged matrices into p. d. ones by diagonal loading and normalization:

$$\mathbf{T}_N^{(0)} = \gamma[\mathbf{R}_N^{ra} + (\boldsymbol{\lambda}_N^{*} + |\boldsymbol{\lambda}_N^{ra}|)\mathbf{I}_N] \tag{28}$$

$$\gamma = \frac{\mathrm{Tr}[\mathbf{R}_N^{ra}]}{\mathrm{Tr}[\mathbf{R}_N^{ra} + (\boldsymbol{\lambda}_N^{*} + |\boldsymbol{\lambda}_N^{ra}|)\mathbf{I}_N]} \tag{29}$$

where $\mathbf{R}_N^{\mathrm{ra}}$ - the redundancy averaged covariance matrix,
$\boldsymbol{\lambda}_N^{\mathrm{ra}}$ - the minimal negative eigenvalue of $\mathbf{R}_N^{\mathrm{ra}}$,
$\boldsymbol{\lambda}_N^{*}$ - the minimal positive eigenvalue, specified by the RMT methodology.

Iterations that start from (28) should not deal with the negative eigenvalues while trimming the positive ones of the matrix $\mathbf{T}_N^{(0)}$. Obviously, the matrix $\mathbf{T}_N^{(0)}$ may be treated as the simplest rectification of the non-p.d. matrix $\mathbf{R}_N^{\mathrm{ra}}$ mentioned above. The main tool of the ToIEP methodology is the Newton's iterative technique that is based on the first order expansion [14]:

$$\boldsymbol{\lambda}_j[\mathbf{T}_N^{(v)} + \boldsymbol{\Delta}_N^{(v+1)}] = \boldsymbol{\lambda}_j[\mathbf{T}_N^{(v)}] + \boldsymbol{\mathcal{U}}_j^{(v)\mathrm{H}}\boldsymbol{\Delta}_N^{(v+1)}\boldsymbol{\mathcal{U}}_j^{(v)} \tag{30}$$

where $\boldsymbol{\mathcal{U}}_j^{(v)}$ is the $j$-th eigenvector of the matrix $\mathbf{T}_N^{(v)}$, and

$$\|\boldsymbol{\Delta}_N^{(v+1)}\| \ll \|\mathbf{T}_N^{(v)}\| \tag{31}$$

Apart from getting a positive definite Toeplitz Hermitian matrix, its proximity to the optimum ML solution will be verified by the sphericity likelihood ratio:

$$\gamma(\widehat{\mathbf{T}}_N|\widehat{\mathbf{R}}_N) = \frac{\det[\widehat{\mathbf{R}}_N\widehat{\mathbf{T}}_N^{-1}]}{[\frac{1}{N}\mathrm{Tr}[\widehat{\mathbf{R}}_N\widehat{\mathbf{T}}_N^{-1}]]^N} \leq 1 \tag{32}$$

where $\widehat{\mathbf{T}}_N$ is the tested matrix. For $\widehat{\mathbf{T}}_N = \mathbf{T}_N$ (true covariance matrix),

$$\gamma(\mathbf{T}_N|\widehat{\mathbf{R}}_N) = \frac{\det[\widehat{\mathbf{S}}_T]}{[\frac{1}{N}\mathrm{Tr}[\widehat{\mathbf{S}}_T]]^N} \tag{33}$$
$$\widehat{\mathbf{S}}_T = \frac{1}{T}\sum_{t=1}^{T}\boldsymbol{\xi}_t\boldsymbol{\xi}_t^{\mathrm{H}}, \quad \boldsymbol{\xi}_t \sim \mathbb{C}\mathcal{N}(0, \mathbf{I}_N).$$

Since

$$\gamma(\widehat{\mathbf{T}}_{\mathrm{ML}}|\widehat{\mathbf{R}}_N) > \gamma(\mathbf{T}_N|\widehat{\mathbf{R}}_N), \tag{34}$$

the estimate $\widehat{\mathbf{T}}_N$, with $\gamma(\widehat{\mathbf{T}}_{\mathrm{ML}}|\widehat{\mathbf{R}}_N)$ consistent with the pdf for $\gamma(\widehat{\mathbf{T}}_N|\widehat{\mathbf{R}}_N)$, is treated "as likely as the true matrix $\mathbf{T}_N$", and therefore treated as the ML-proxy estimate. If

$$\gamma(\widehat{\mathbf{T}}_N|\widehat{\mathbf{R}}_N) \ll \gamma(\mathbf{T}_N|\widehat{\mathbf{R}}_N)\,, \tag{35}$$

i.e. the value $\gamma(\widehat{\mathbf{T}}_N|\widehat{\mathbf{R}}_N)$ is below the support of the pdf for $\gamma(\mathbf{T}_N|\widehat{\mathbf{R}}_N)$, then the estimate $\widehat{\mathbf{T}}_N$ is treated as being far from the ML optimum [9]. For "spiked" covariance matrices $\mathbf{T}_N$ with $m$ last eigenvalues equal to the white noise additive power, we may use the "spiky" sphericity test (20). Once again, $\widehat{\mathbf{T}}_N$ may be treated as appropriate if $\gamma_m(\widehat{\mathbf{T}}_m|\widehat{\mathbf{R}}_N)$ is within the support of the pdf of $\gamma_m(\mathbf{T}_m|\widehat{\mathbf{R}}_N)$. Let us now specify the redundancy averaged matrix rectification procedure.

### *A. Redundancy Matrix Rectification*

*1) Noise Subspace Dimension Estimate:* For the signal subspace $m$ estimation we are using the sphericity test calculated for the matrix

$$\widehat{\mathbf{R}}_m = \boldsymbol{\mathcal{U}}_{N-m}\dot{\boldsymbol{\Lambda}}_{N-m}\boldsymbol{\mathcal{U}}_{N-m}^{\mathrm{H}} + \dot{\boldsymbol{\lambda}}_{m+1}\boldsymbol{\mathcal{U}}_m\boldsymbol{\mathcal{U}}_m^{\mathrm{H}} \tag{36}$$

so that

$$\mathrm{LR}(\widehat{\mathbf{R}}_m|\widehat{\mathbf{R}}_N) = \frac{\prod_{j=N-m+1}^{N}\hat{\boldsymbol{\lambda}}_{\mathrm{j}}}{[\frac{1}{m}\sum_{j=N-m+1}^{N}\hat{\boldsymbol{\lambda}}_{\mathrm{j}}]^{N-m}} \tag{37}$$

and compare it against the support of the pdf of $\gamma(\mathbf{T}_N|\widehat{\mathbf{R}}_N)$. The noise subspace $m$ dimension is accepted if $\mathrm{LR}(\widehat{\mathbf{R}}_m|\widehat{\mathbf{R}}_N)$ is within the support of the pdf of $\gamma(\mathbf{T}_N|\widehat{\mathbf{R}}_N)$.

*2) Modifying Signal Subspace ("Spiky") Eigenvalues Estimates By The RMT Methodology:* According to Theorem 3 in [11], the following quantities are strongly ($T$, $N$) consistent:

$$\gamma_n = \frac{T}{K_n}\sum_{K\in\varkappa_n}(\hat{\boldsymbol{\lambda}}_k - \hat{\boldsymbol{\mu}}_k) \tag{38}$$

Here $\hat{\boldsymbol{\lambda}}_k$ is the traditional $k$-th eigenvalue with multiplicity $K_n$ and $\hat{\mu}_1 < \mu_2 < \cdots < \mu_N$ are the real-valued solutions to the following problem:

Find all $\mu \in \mathbf{R}$ such that

$$\frac{1}{N}\sum_{K=1}^{N}\frac{\hat{\boldsymbol{\lambda}}_k}{\hat{\boldsymbol{\lambda}}_k-\mu}=\frac{N}{T}\ (<1) \tag{39}$$

$$\varkappa_m=\left\{\sum_{r=1}^{m-1}K_r+1,\sum_{r=1}^{m-1}K_r+2,\dots,\sum_{r=1}^{m}K_r\right\}, \tag{40}$$

$K_m$ is the multiplicity of the eigenvalue $\gamma_m$

As demonstrated in [11], modification of the sample eigenvalues $\hat{\boldsymbol{\lambda}}_n$ into the eigenvalues $\boldsymbol{\gamma}_n$, brings $\boldsymbol{\gamma}_n$ much closer to the true eigenvalues of the covariance matrix $\mathbf{T}_N$, even for the modest $N$ and $T$ values.

*3) Alternating Reconstruction of a P.D. Hermitian Toeplitz Matrix with the Given Eigenvectors*

*a) Phases Adjustment:* As mentioned above, the covariance lag's phases get iteratively trimmed by the small phase increments that allow for first order expansion accuracy:

$$\begin{aligned}\exp\left(i\left(\boldsymbol{\psi}_n^{(v)}+\boldsymbol{\delta}_n^{(v+1)}\right)\right)\\ &=\exp\left(i\boldsymbol{\psi}_n^{(v)}\right)(1+i\sin(\boldsymbol{\delta}_n^{(v+1)})))\\ &\approx\exp(i\boldsymbol{\psi}_n^{(v)})(1+i\,\boldsymbol{\delta}_n^{(v+1)})\end{aligned} \tag{41}$$

and

$$\boldsymbol{\lambda}_j[Z+dZ]=\boldsymbol{\lambda}_j[Z]+\boldsymbol{\mathcal{U}}_j^{\mathrm{H}}dZ\boldsymbol{\mathcal{U}}_j \tag{42}$$

where $\boldsymbol{\mathcal{U}}_j$ is the $j$-th eigenvector of the matrix $\mathbf{Z}$. Therefore, the increments $\boldsymbol{\delta}_n^{(v+1)}$ should be kept small to keep these first-order expansions (41), (42) accurate. With respect to (41), we have

$$\begin{aligned}\mathbf{T}_N^{(v+1)}&\equiv\mathbf{T}_N\left(\boldsymbol{\psi}_{N-1}^{(v+1)}\right)\\ &=\mathbf{T}_N\left(\boldsymbol{\psi}_{N-1}^{(v)}\right)+i\sum_{n=1}^{N-1}t_n\,\mathbf{A}\left(\boldsymbol{\psi}_n^{(v)}\right)\delta_n^{(v+1)},\end{aligned} \tag{43}$$

where $t_n>0$ is the covariance lag modulus, unchanged during these iterations, and

$$\mathbf{A}\left(\boldsymbol{\psi}_n^{(v)}\right)=\begin{bmatrix}0 & \dots & 0 & e^{j\psi_n^{(v)}} & 0 & & 0\\ \vdots & & & & e^{j\psi_n^{(v)}} & & \\ 0 & & & & & & \\ -e^{-j\psi_n^{(v)}} & & & & & \ddots & \\ 0 & -e^{-j\psi_n^{(v)}} & & & & & e^{j\psi_n^{(v)}}\\ & & \ddots & & & & \\ 0 & \dots & 0 & -e^{-j\psi_n^{(v)}} & 0 & \dots & 0\end{bmatrix}. \tag{44}$$

In (44) $\mathbf{A}(\boldsymbol{\psi}_n^{(v)})$ is a skew Hermitian matrix and therefore the matrix $i\mathbf{A}(\boldsymbol{\psi}_n^{(v)})$ is a Hermitian matrix [15]. For small $\boldsymbol{\delta}_{N-1}^{(v+1)}$, using the first order expansion (42), we get:

$$\boldsymbol{\Lambda}_N^{(v+1)}=\boldsymbol{\Lambda}_N^{(v)}+\mathbf{B}_{N,N-1}^{(v)}\boldsymbol{\delta}_{N-1}^{(v+1)} \tag{45}$$

where

$$\begin{gathered}\mathbf{B}_{N,N-1}^{(v)}=\left[t_n\boldsymbol{\mathcal{U}}_m^{(\mathrm{v})\mathrm{H}}i\mathbf{A}\left(\boldsymbol{\psi}_n^{(v)}\right)\boldsymbol{\mathcal{U}}_m^{(\mathrm{v})}\right]\\ m=0,\dots,N-1\\ n=1,\dots,N-1\end{gathered} \tag{46}$$

Therefore, we get

$$\boldsymbol{\delta}_{N-1}^{(v+1)}=c[\mathbf{B}_{N,N-1}^{(v)\mathrm{H}}\mathbf{B}_{N,N-1}^{(v)}]^{-1}\mathbf{B}_{N,N-1}^{(v)\mathrm{H}}\boldsymbol{\Delta}_N^{(v)} \tag{47}$$

$$\boldsymbol{\Delta}_N^{(v)}=\boldsymbol{\Lambda}_N^{(v)}-\dot{\boldsymbol{\Lambda}}_N \tag{48}$$

where

$\dot{\boldsymbol{\Lambda}}_N$ - "target" eigenvalues derived by the RMT technology in (38)-(40)

$c$ - "scaling" constant that keeps $\boldsymbol{\delta}_N^{(v+1)}$ small for meeting the first-order expansions (41), (43).

*b) Moduli Adjustment:* When the Newton's procedure (47), (48) converges to a solution that's still far away from the desired optimum, we may fix the optimized phases and iterate over ($N$ - 1) moduli of the covariance lags. The Toeplitz matrix innovation for the moduli upgrade is

$$\mathbf{T}_N^{(v+1)}\equiv\mathbf{T}_N\left(\mathbf{P}_{N-1}^{(v+1)}\right)=\mathbf{T}_N^{(v)}+\sum_{n=1}^{N-1}\dot{\mathbf{A}}\,(\boldsymbol{\psi}_n)\delta_n^{(v+1)}, \tag{49}$$

where

$$\mathbf{A}(\boldsymbol{\psi}_n)=\begin{bmatrix}0 & \dots & 0 & e^{j\psi_n} & 0 & & 0\\ \vdots & & & & e^{j\psi_n} & & \\ 0 & & & & & & \\ e^{-j\psi_n} & & & & & \ddots & \\ 0 & e^{-j\psi_n} & & & & & e^{j\psi_n}\\ & & \ddots & & & & \\ 0 & \dots & 0 & e^{-j\psi_n} & 0 & \dots & 0\end{bmatrix}. \tag{50}$$

This representation leads to the equation

$$\boldsymbol{\Lambda}_N^{(v+1)}=\boldsymbol{\Lambda}_N^{(v)}+\dot{\mathbf{B}}_{N,N-1}^{(v)}\boldsymbol{\delta}_{N-1}^{(v+1)} \tag{51}$$

where

$$\begin{gathered}\dot{\mathbf{B}}_{N,N-1}^{(v)}=\left[\boldsymbol{\mathcal{U}}_n^{(\mathrm{v})\mathrm{H}}\dot{\mathbf{A}}(\boldsymbol{\psi}_m)\boldsymbol{\mathcal{U}}_n^{(\mathrm{v})}\right]\\ m=1,\dots,N-1\\ n=0,\dots,N-1\end{gathered} \tag{52}$$

and to the iterative procedure

$$\boldsymbol{\delta}_{N-1}^{(v+1)}=c[\dot{\mathbf{B}}_{N,N-1}^{(v)\mathrm{H}}\dot{\mathbf{B}}_{N,N-1}^{(v)}]^{-1}\dot{\mathbf{B}}_{N,N-1}^{(v)\mathrm{H}}\boldsymbol{\Delta}_{N-1}^{(v)} \tag{53}$$

with the step-size $c$ that keeps $\boldsymbol{\delta}_{N-1}^{(v+1)}$ small enough for the first-order expansion to be accurate.

If the specified "RMT" eigenvalues $\boldsymbol{\Lambda}_N^*$ are not reached by the moduli adjustments, we then repeat our iterations over phases and alternate between the phases and moduli of the covariance lags iterative adjustment until the desired solution is achieved. Let us repeat that convergence to the specified eigenvalues does not guarantee that this solution has a sufficiently high likelihood ratio. The possibility of even "walking" further away from the desired solution with a high likelihood while pursuing the specified eigenvalues cannot be completely dismissed. Simulations should reveal the actual nature of the convergence in the iterative procedures, leading closer to the prescribed eigenvalues.

*B. Alternative Method of Transition from the P.D. Hermitian Sample Matrix to a P.D. Hermitian Toeplitz Matrix*

The iterative routines introduced above that "rectify" the redundancy averaged Toeplitz matrix are not the only possibility to transform a p.d. Hermitian sample matrix into a p.d. Hermitian Toeplitz matrix. In [24], we introduced a method of a p.d. Hermitian Toeplitz matrix reconstruction, that has the same Maximum Entropy spectrum as the Hermitian sample matrix. Unfortunately, due to computational complexity it was not broadly used, but with current computational advances it could be treated as an operational alternative to the discussed above techniques. This method is based on the Lemma proven in [24].

Lemma: In order for an arbitrary $N$-dimensional vector $\mathbf{P}$ with $p_1 > 0$ to be represented as

$$\mathbf{P} = \mathbf{T}_N^{-1}\boldsymbol{e}_1 \quad (54)$$

where $\mathbf{T}_N$ is a p.d. Hermitian Toeplitz matrix and $\mathbf{e}_1^{\mathrm{T}} = (1,0,\dots,0)$, it is necessary and sufficient that the polynomial

$$\mathbf{P}(\mathrm{z}) = \sum_{n=1}^{N} \mathbf{P}_n\, \mathbf{z}^{n-1} \quad (55)$$

have no zeros inside the unit disk:

$$\mathbf{P}(\mathrm{z}) \neq 0, \;\; \text{for } |z| = 1 \quad (56)$$

Given this Lemma, the problem of the p.d. Toeplitz matrix reconstruction given the vector $\widehat{\mathbf{W}}$

$$\widehat{\mathbf{W}} = \frac{\widehat{\mathbf{R}}_N^{-1}\boldsymbol{e}_1}{\mathbf{e}_1^{\mathrm{T}}\widehat{\mathbf{R}}_N^{-1}\boldsymbol{e}_1} \quad (57)$$

is to reconstruct the vector $\mathbf{P}$ with the property (56) such that

$$\left|\widehat{\mathbf{W}}(z)\right| = |\mathbf{P}(z)|, \quad \text{for } |z| = 1, p_1 \geq 1 \quad (58)$$

In general, the vector $\widehat{\mathbf{W}}$ (57) may fail to satisfy and have a certain number $m$ ($m < N$) of zeros inside the unit disk $|z| = 1$.

For this case, we use the very well-known representation of an arbitrary polynomial $\widehat{\mathbf{W}}(z)$ of the form:

$$\widehat{\mathbf{W}}(z) \; = b(z)\widehat{\mathbf{P}}(z) \quad (59)$$

where $b(z)$ is a product of elementary factors

$$b(z) = \prod_{n=1}^{m} \frac{z - z_n}{1 - z_n^* z}, \quad 0 < |z| < 1 \quad (60)$$

constructed by $m$ zeros $z_n$ of the polynomial $\widehat{\mathbf{W}}(z)$ located within the unit disk $|z| = 1$, taking multiplicity into account. In (59) $\widehat{\mathbf{P}}(z)$, is the polynomial of the same degree as $\widehat{\mathbf{W}}(z)$, but with no zeros inside the unit disk $|z| = 1$. With the additional condition $\mathbf{P}_1 = \mathbf{P}(0) > 0$, we have the unique representation:

$$\widehat{\mathbf{P}}(z) = e^{i\gamma} \prod_{n=1}^{m} \frac{1 - z_n^* z}{z - z_n} \widehat{\mathbf{W}}(z) \quad (61)$$

$$\gamma = \pi m + \sum_{n=1}^{m} \arg(z_n)\,. \quad (62)$$

Due to the properties of this product, we have:

$$\left|\widehat{\mathbf{P}}(z)\right| = \left|\widehat{\mathbf{W}}(z)\right|, \quad \text{for } |z| = 1 \quad (63)$$

and

$$\widehat{\mathbf{P}} = \mathbf{P}(0) = \prod_{n=1}^{m} |z_n|^{-1} > 1 \quad (64)$$

since all of the individual roots $z_n$ in (63) lie inside the unit disk $|z| = 1$.

Now, by using the Gohberg-Semencul theorem [28] we get:

$$\widehat{\mathbf{P}}_1\mathbf{T}_N^{-1} = \begin{bmatrix} \widehat{\mathbf{P}}_1 & 0 & \cdots & 0 \\ \widehat{\mathbf{P}}_2 & \widehat{\mathbf{P}}_1 & \cdots & 0 \\ \cdots & \cdots & \ddots & 0 \\ \widehat{\mathbf{P}}_N & \widehat{\mathbf{P}}_{N-1} & \cdots & \widehat{\mathbf{P}}_1 \end{bmatrix} \begin{bmatrix} \widehat{\mathbf{P}}_1^* & \widehat{\mathbf{P}}_2 & \cdots & \mathbf{P}_N^* \\ 0 & \widehat{\mathbf{P}}_1^* & \cdots & \mathbf{P}_{N-1}^* \\ & & \ddots & \\ 0 & 0 & \cdots & \widehat{\mathbf{P}}_1^* \end{bmatrix} - \begin{bmatrix} 0 & 0 & & \cdots & 0 \\ \widehat{\mathbf{P}}_N^* & & & & \vdots \\ & \widehat{\mathbf{P}}_N^* & & & \vdots \\ & & \ddots & & \vdots \\ \widehat{\mathbf{P}}_2^* & \widehat{\mathbf{P}}_3^* & & \widehat{\mathbf{P}}_N^* & 0 \end{bmatrix} \begin{bmatrix} 0 & \mathbf{P}_N & \cdots & \mathbf{P}_2 \\ & & & \vdots \\ & & \ddots & \mathbf{P}_N \\ 0 & 0 & \cdots & 0 \end{bmatrix} \quad (65)$$

Thus, the inverse of the p.d. Toeplitz matrix may be reconstructed using the vector $\hat{\mathbf{R}}_N^{-1}\boldsymbol{e}_1/(\mathbf{e}_1^{\mathrm{T}}\hat{\mathbf{R}}_N^{-1}\boldsymbol{e}_1)$ by finding the roots of the polynomial $\hat{\mathbf{W}}(z)$. This procedure generates a p.d. Toeplitz matrix with the same Maximum Entropy spectrum $\hat{\mathbf{W}}^{-1}(z), |z| = 1$ as the sample Hermitian matrix $\hat{\mathbf{R}}_N$. While the ME spectrum reflects on the main properties of a p.d. Toeplitz matrix, the eigenvalues of the reconstructed matrix may be off from the ones recommended by RMT (38)-(40). Obviously, the transformation introduced above of a Hermitian p.d. matrix into a p.d. Toeplitz Hermitian matrix, may be used in the alternating projection routine with the replacement of the eigenvalues of the restored Toeplitz matrix by the RMT specified ones, followed by transformation of the obtained Hermitian matrix into the p.d. Toeplitz one. Progression of the likelihood ratio associated with these alternating projections is analyzed in the next section.

## IV. Simulation Results

We start this section with the analysis of the redundancy averaged sample matrix rectification efficiency by the proposed alternating iterative Newton's technique from the ToIEP methodology. Recall that the ToIEP technique for symmetric Toeplitz matrix with the very limited number of solutions to the problem of matrix reconstruction, given its elements' moduli and eigenvalues, was extended by us over the class of Hermitian Toeplitz matrices in anticipation that by starting from the redundancy averaged matrix, this rectification should lead sufficiently close to the desired optimum solution. The simulations should validate this anticipation. For analysis we selected the same Hermitian Toeplitz matrix (8)-(10) with the sample volume $T = 85\,(T/N = 5)$. Recall that in our alternating optimization routine, we first update the phases of the redundancy averaged covariance lags iteratively, leaving their moduli unchanged. After conversion of the phase optimizations, we switch to the iterative moduli upgrade, keeping the phases unchanged, and then continue alternating.

Here we report the results for $5{\cdot}10^3$ total iterations. First, in Fig. 21 we observe the very slow convergence of the optimized eigenvalues, from $\|\boldsymbol{\Lambda}_N(1) - \boldsymbol{\Lambda}_N^*\|_2 = 1.15$ at the first iteration, to $\|\boldsymbol{\Lambda}_N(5000) - \boldsymbol{\Lambda}_N^*\|_2 = 0.72$ at the final iteration.

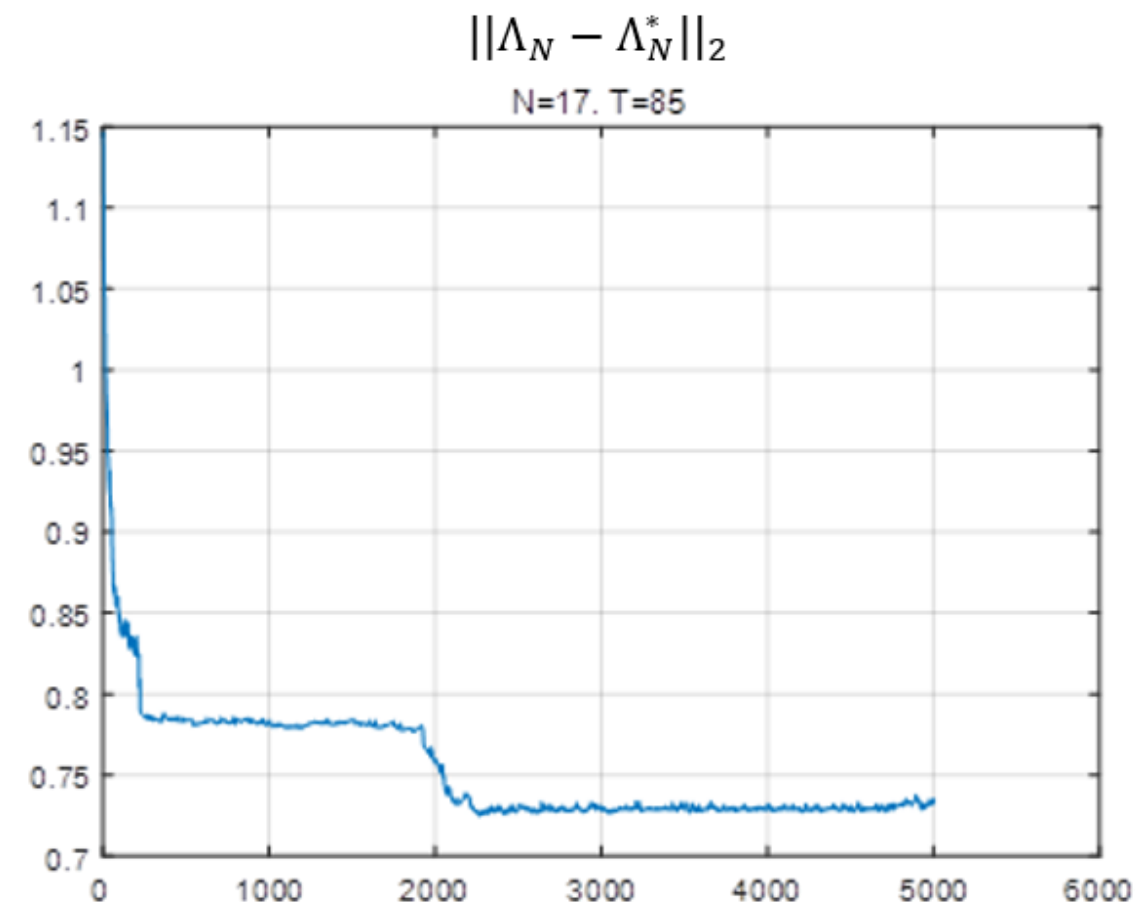

Fig. 21. Convergence of optimized to desired eigenvalues. $\Lambda_N$ = optimized eigenvalues. $\Lambda_N^*$ = target eigenvalues (RMT)

As discussed, this slow eigenvalue convergence is due to the very small step-size that keeps the first-order expansion accurate. The actual eigenvalues are introduced in the TABLE III. for:

1) $\lambda(\mathbf{T_N})$ - the actual (true) covariance matrix $\mathbf{T}_N$
2) $\lambda(\hat{\mathbf{R}})$ - the redundancy averaged Toeplitz matrix with negative eigenvalues
3) $\lambda(\hat{\mathbf{T}}_N^{\mathrm{ra}})$ - the diagonally loaded redundancy averaged matrix with all positive eigenvalues used to initialize the iterative refinement
4) $\lambda^*$ - "target" eigenvalues, calculated using the RMT methodology
5) $\lambda(\hat{\mathbf{T}}_N)$ (phase trim) - eigenvalues after first "phase-only" iterations
6) $\lambda(\hat{\mathbf{T}}_N)$ (phase & moduli trim) - eigenvalues after alternating phases and moduli optimization at $v$ = 5000

TABLE III.

| $\lambda(T_N)$ | $\lambda(\hat{R})$ | $\lambda(\hat{T}_N^{ra})$ | $\lambda^*$ |
|---|---|---|---|
| 1.4964 | 1.992888 | 2.314267 | 1.7082 |
| 1.4248 | 1.533716 | 1.348330 | 1.3437 |
| 1.1368 | 1.336132 | 1.151430 | 1.3298 |
| 1.0009 | 1.121411 | 0.943905 | 0.98154 |
| 1.0001 | 0.999965 | 0.788519 | 1.1373 |
| 0.99237 | 0.833157 | 0.612460 | 0.92644 |
| 0.81499 | 0.709042 | 0.518154 | 0.92292 |
| 0.4506 | 0.314187 | 0.480149 | 0.36883 |
| 0.15841 | 0.139769 | 0.399148 | 0.16918 |
| 0.023622 | 0.015833 | 0.382318 | 0.017623 |
| 0.0020411 | 0.001712 | 0.295009 | 0.001923 |
| 0.00020973 | 0.000233 | 0.225881 | 0.000117 |
| 0.00010417 | 0.000102 | 0.074540 | 0.000117 |
| 0.00010011 | 0.000086 | -0.062338 | 0.000117 |

| $\lambda(T_N)$ | $\lambda(\widehat{R})$ | $\lambda(\widehat{T}_N^{ra})$ | $\lambda^*$ |
|---|---|---|---|
| 0.0001 | 0.000066 | -0.086109 | 0.000117 |
| 0.0001 | 0.000059 | -0.142516 | 0.000117 |
| 0.0001 | 0.000051 | -0.244740 | 0.000117 |

| $\lambda(\widehat{T}_N)$ (*phase trim*) | $\lambda(\widehat{T}_N)$ (*phase* & *moduli trim*) |
|---|---|
| 1.3566 | 1.7044 |
| 1.1331 | 1.3507 |
| 1.0877 | 1.3226 |
| 0.93154 | 1.1081 |
| 0.89627 | 1.0132 |
| 0.83446 | 0.94448 |
| 0.73708 | 0.89763 |
| 0.44579 | 0.38676 |
| 0.29105 | 0.16225 |
| 0.23937 | 0.061469 |
| 0.1921 | 0.016353 |
| 0.17445 | 0.014859 |
| 0.15533 | 0.0087405 |
| 0.14268 | 0.0070984 |
| 0.11835 | 0.0053186 |
| 0.10276 | 0.0030615 |
| 0.069587 | 0.0010855 |

As follows from the presented data with 5000 iterations, we did not reach the RMT-provided "goal" eigenvalues, though we noticeably reduced the difference between the optimized and the desired covariance matrix eigenspectrum. Yet, analysis of the "spiked" sphericity test (Fig. 22), the regular sphericity test (Fig. 23), and the spectral norm of the “error matrix" (Fig. 24), tells us a different story.

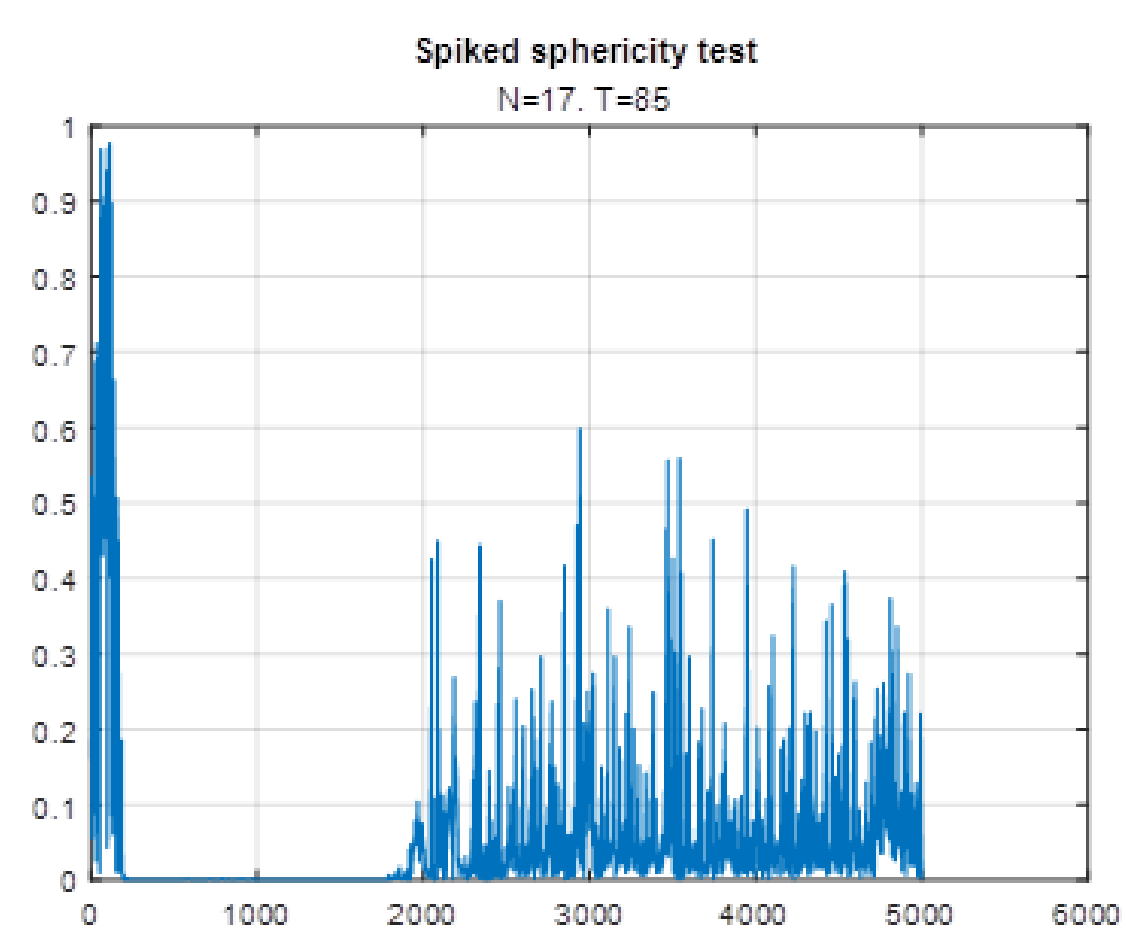

Fig. 22. Spiked sphericity test

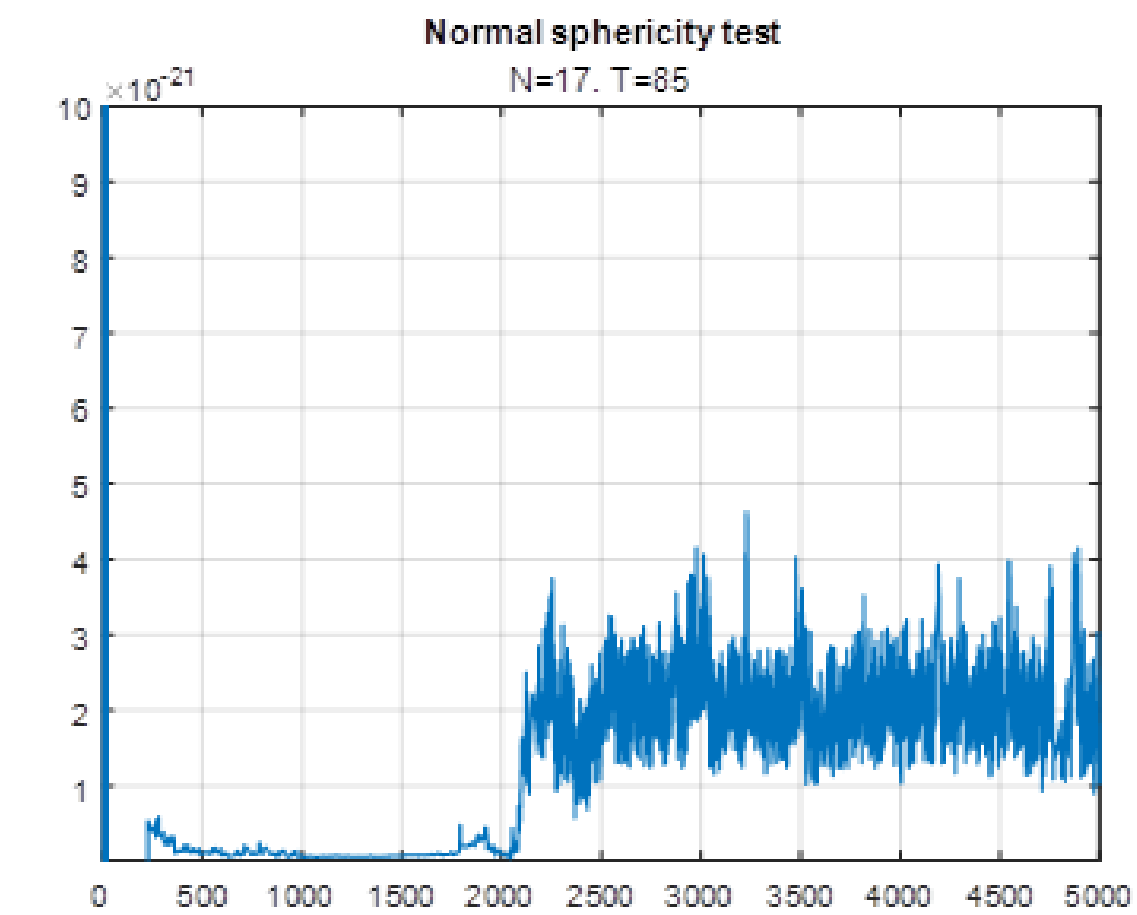

Fig. 23. Normal sphericity test

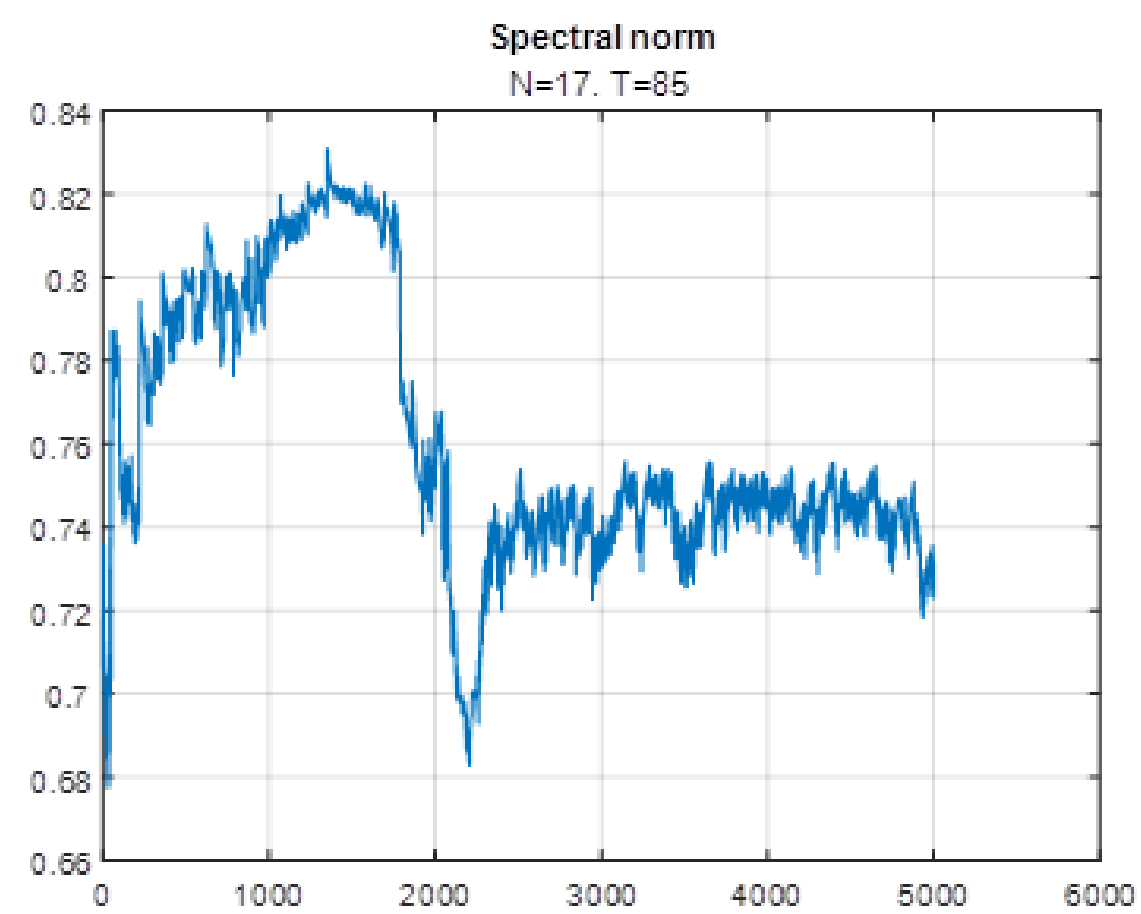

Fig. 24. Spectral norm

Indeed, in terms of these criteria, the best results for all three norms were achieved after ~80-90 iterations, followed by the rapid decline for all three criteria. For example, the "spiked" sphericity test dropped to almost zero. After ~2200 iterations these criteria somewhat improved, but never got back to the levels observed after 80-90 iterations.

Therefore, the conducted trials demonstrated that our anticipation regarding starting from the "close to the optimum” redundancy averaged solution failed and the optimization, in fact, moved the matrix further away from the desired optimum. This result made it clear that by starting from the "close to the optimum” solution, we still could converge to a completely inappropriate Hermitian Toeplitz matrix, yet with the eigenvalues close to the desired ones. For this reason, we moved to the alternative technique described above, that allows for direct reconstruction of the p.d. Toeplitz matrix from the sample Hermitian p.d. matrix. First, we analyze the solution we get by the transformation, described in Section II., sub-section B, of the sample matrix with $T$ = 51, 85, 170 and 340. The same true covariance matrix (8)-(10) was used in these simulations. For $n = 10^2$ Monte Carlo trials, in Fig. 25-Fig. 40 we plotted the sample pdf's of the "spiky" sphericity test and the regular sphericity test for all the above mentioned sample volumes.

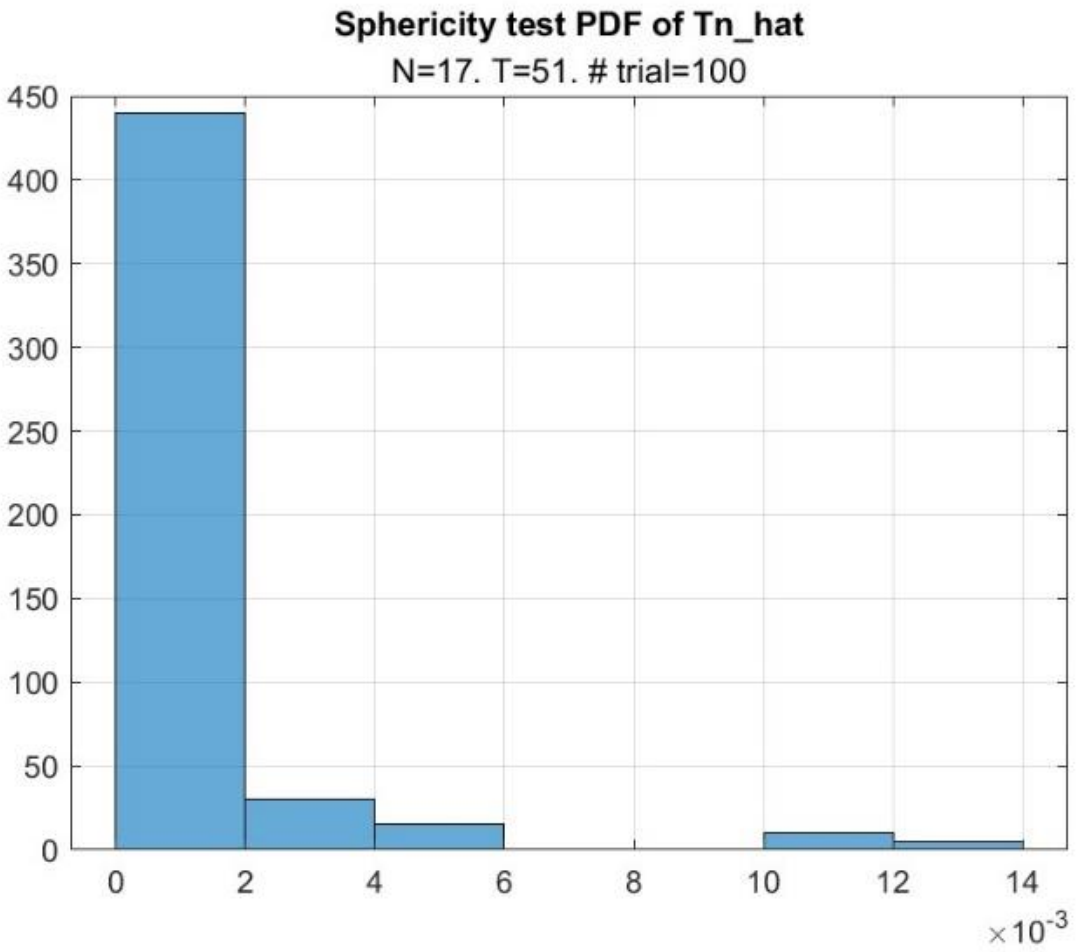

Fig. 25. Normal sphericity test PDF of $\widehat{\boldsymbol{T}}_N$. N=17. T=51.

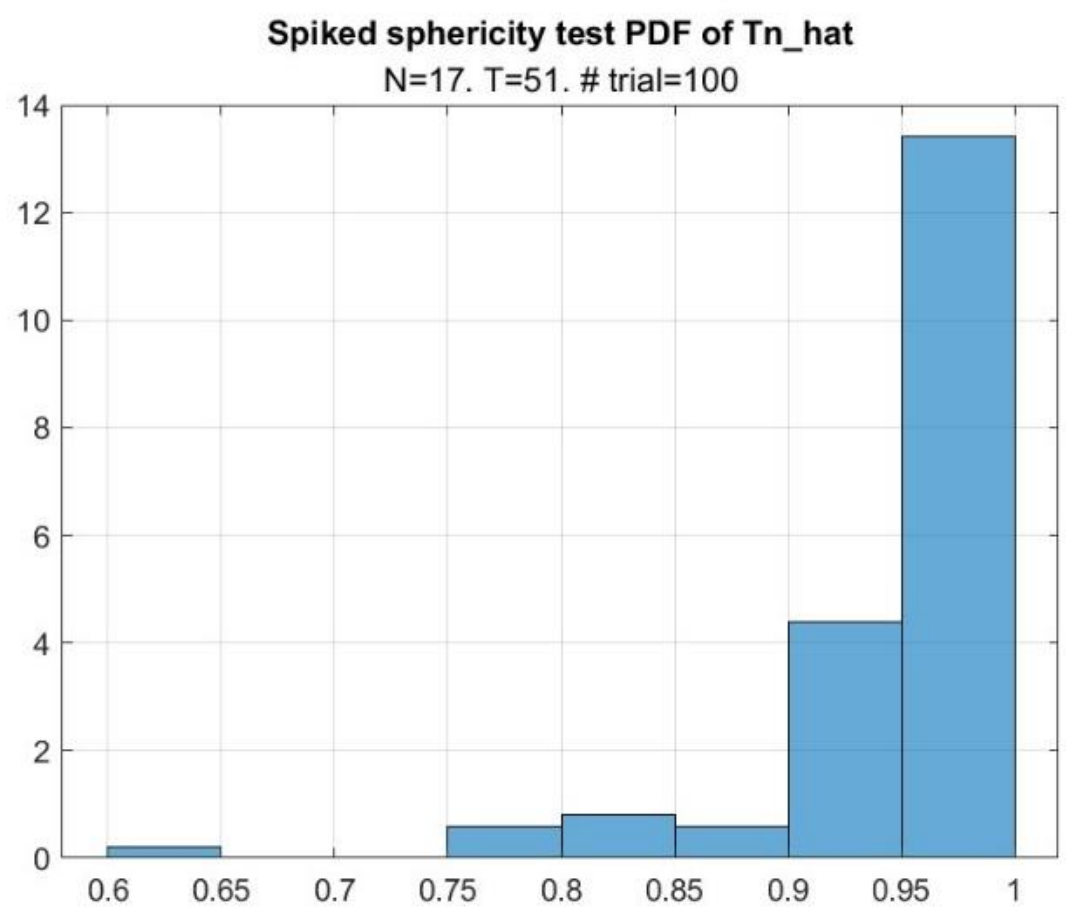

Fig. 26. Spiked sphericity test PDF of $\widehat{\boldsymbol{T}}_N$. N=17. T=51.

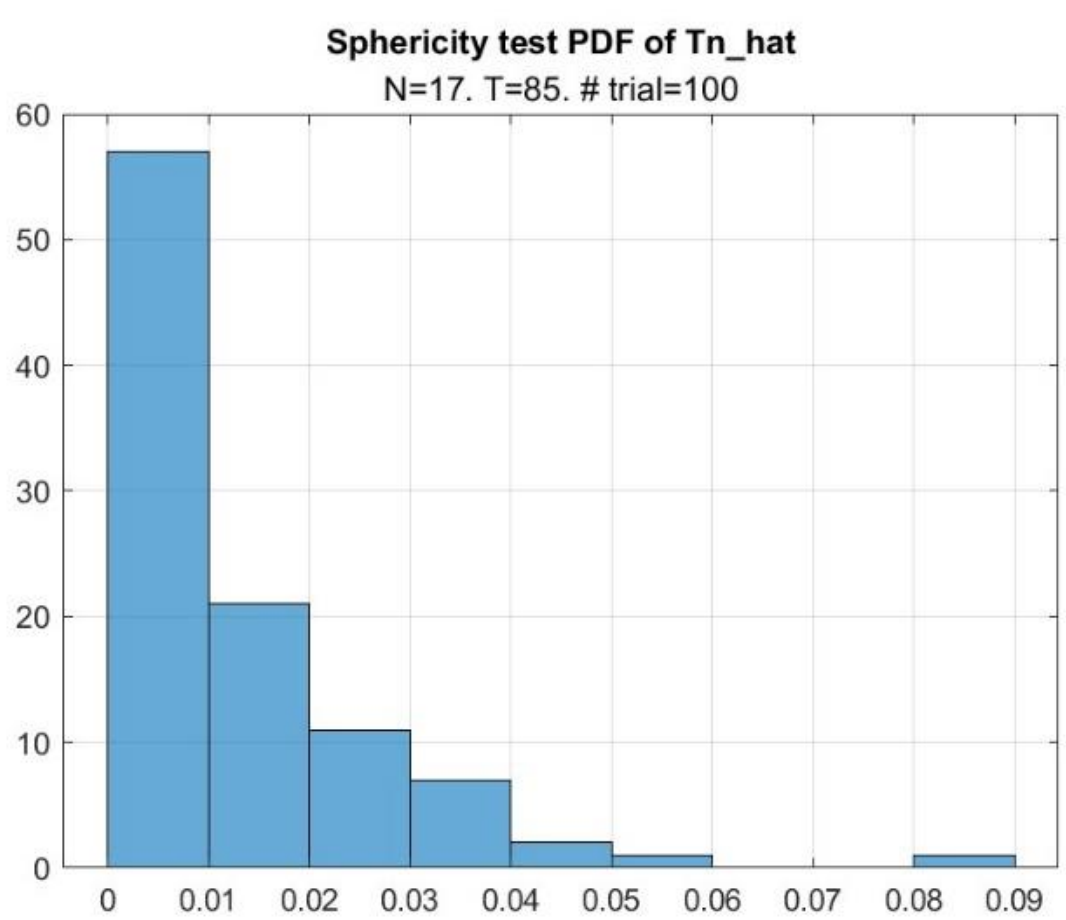

Fig. 27. Normal sphericity test PDF of $\widehat{\boldsymbol{T}}_N$. N=17. T=85.

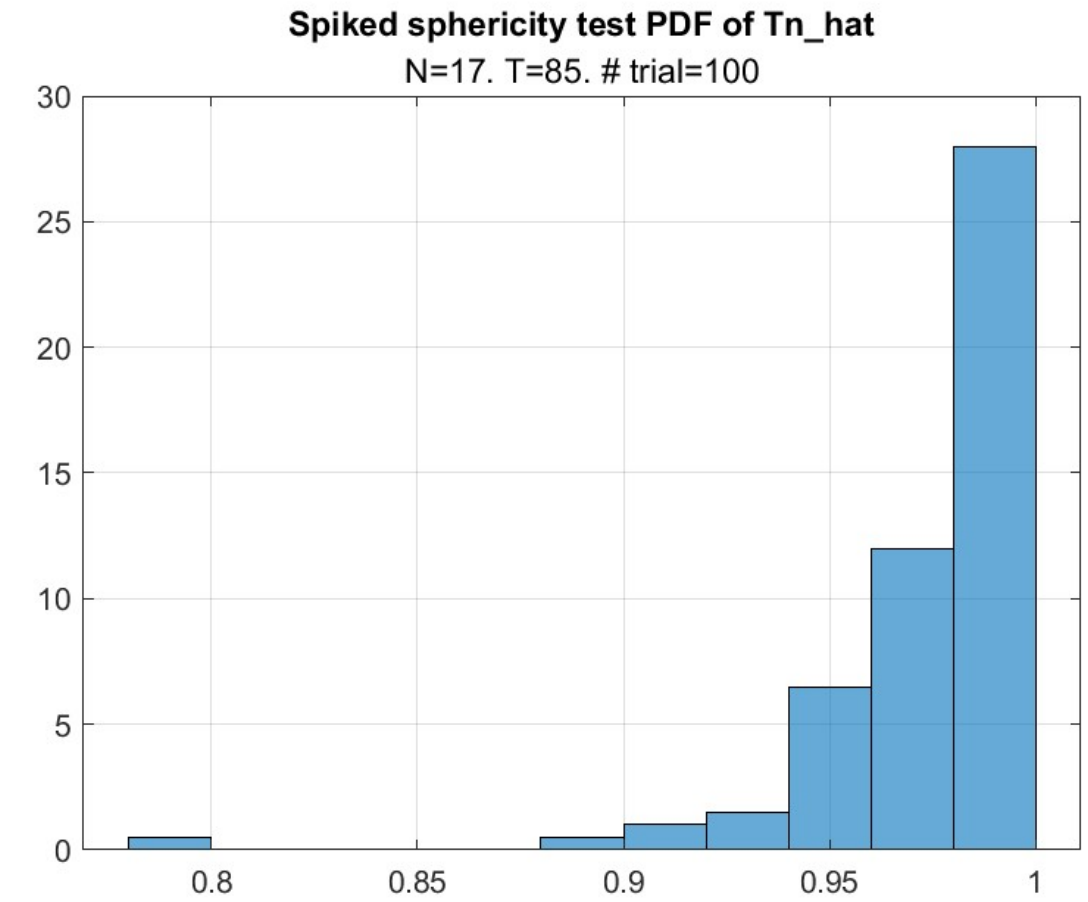

Fig. 28. Spiked sphericity test PDF of $\widehat{\boldsymbol{T}}_N$. N=17. T=85.

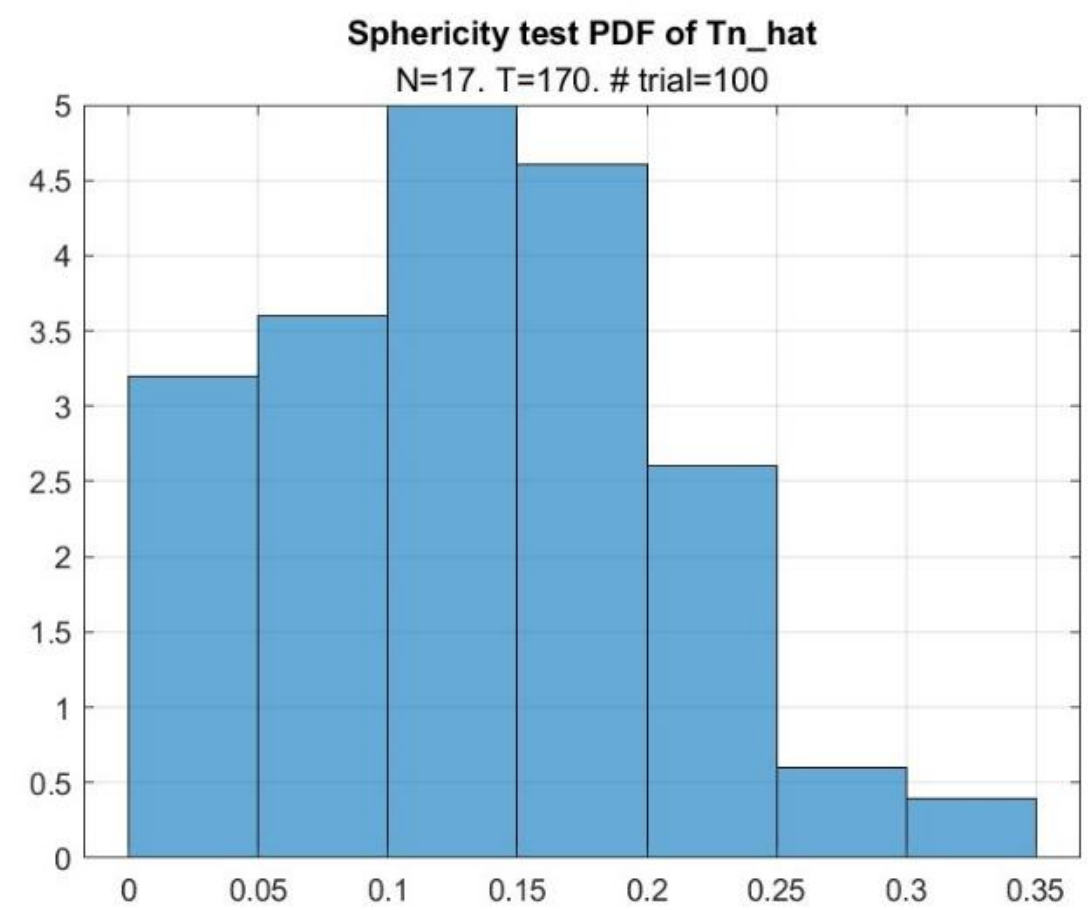

Fig. 29. Normal sphericity test PDF of $\widehat{\boldsymbol{T}}_N$. N=17. T=170.

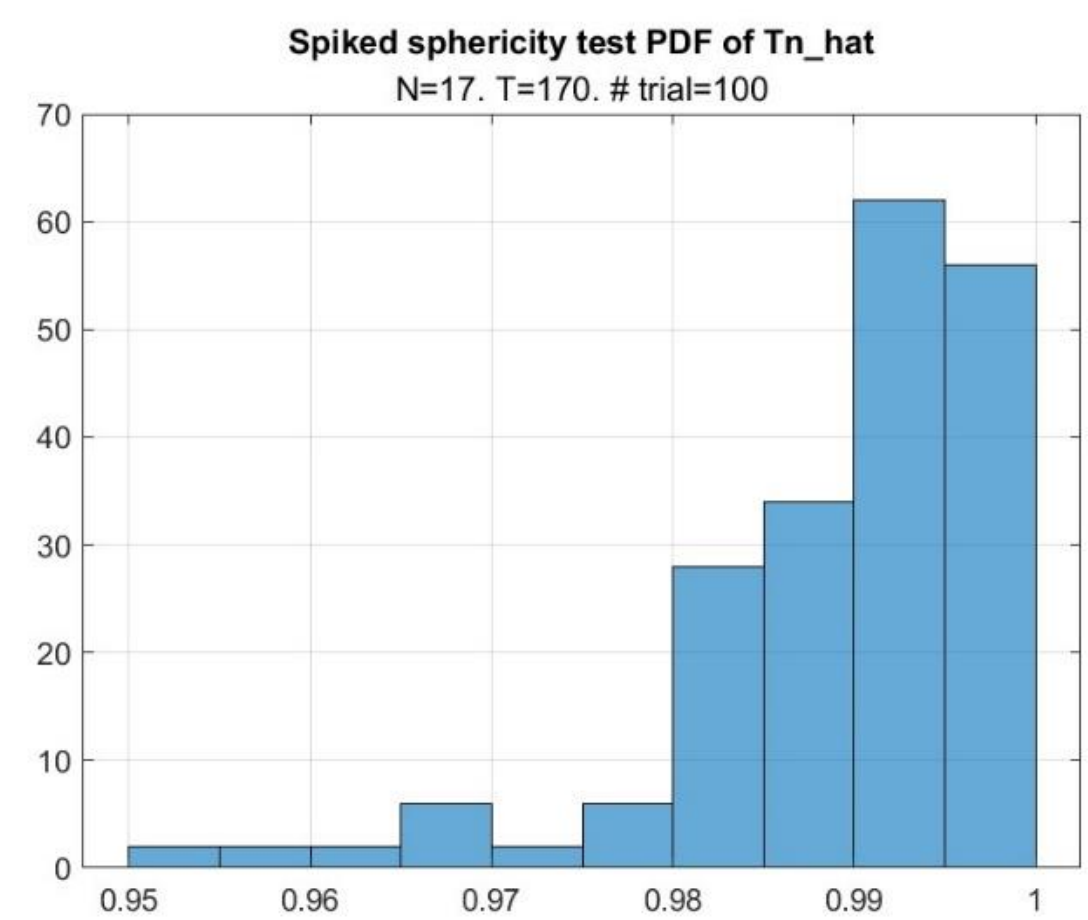

Fig. 30. Spiked sphericity test PDF of $\widehat{\boldsymbol{T}}_N$. N=17. T=170.

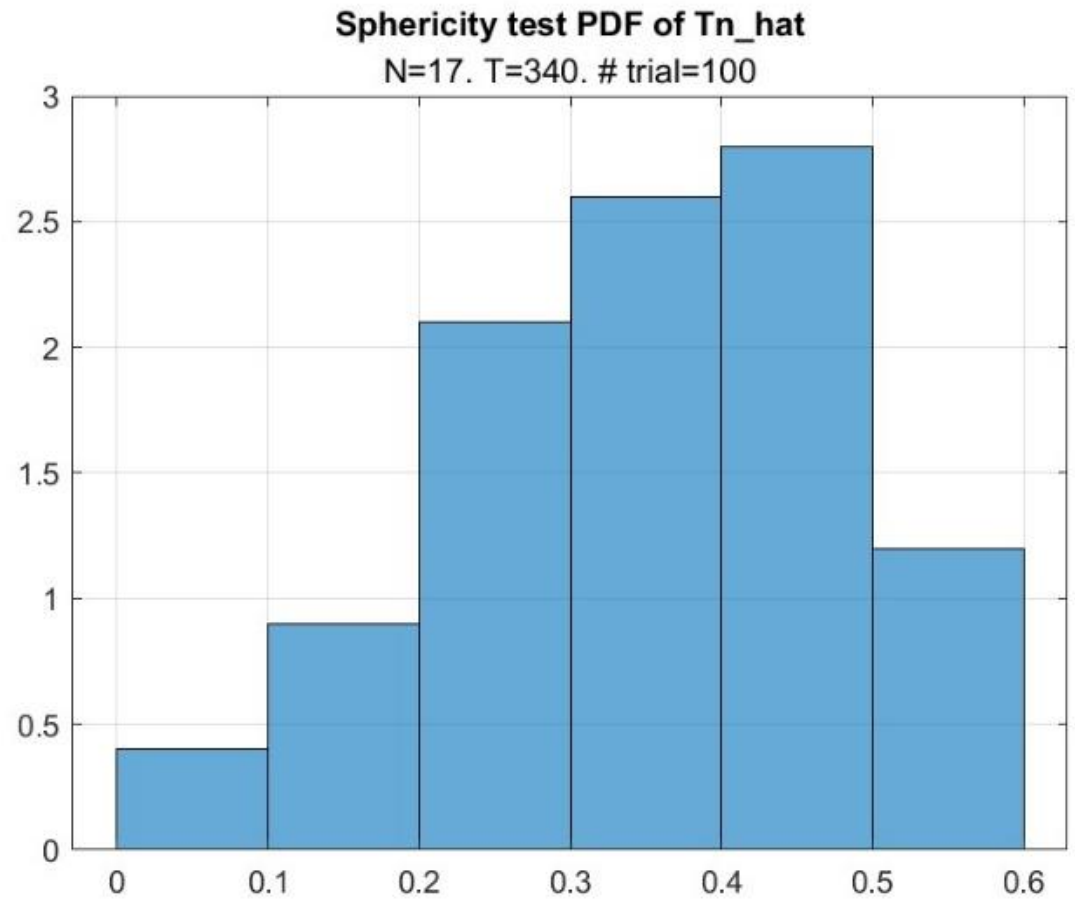

Fig. 31. Normal sphericity test PDF of $\widehat{\boldsymbol{T}}_N$. N=17. T=340.

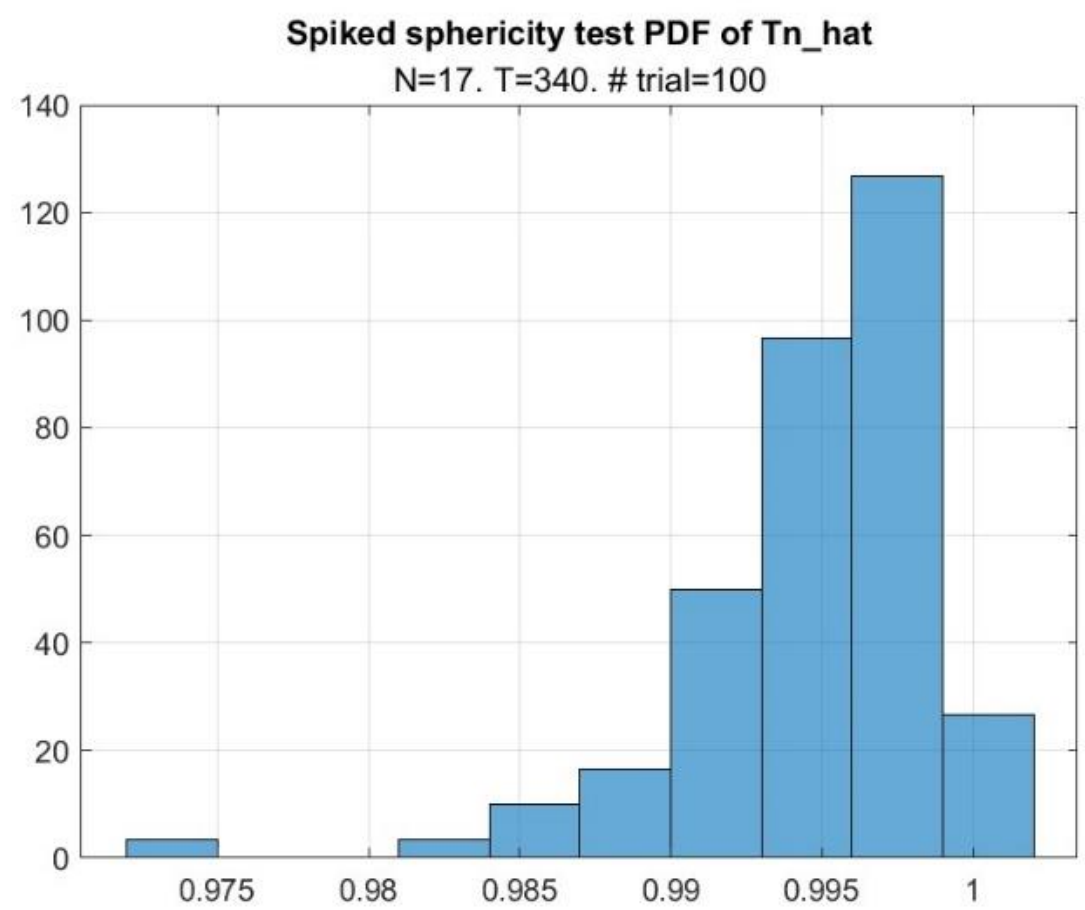

Fig. 32. Spiked sphericity test PDF of $\widehat{\boldsymbol{T}}_N$. N=17. T=340.

Similar pdf's were calculated for the true covariance matrix $\mathbf{T}_N$.

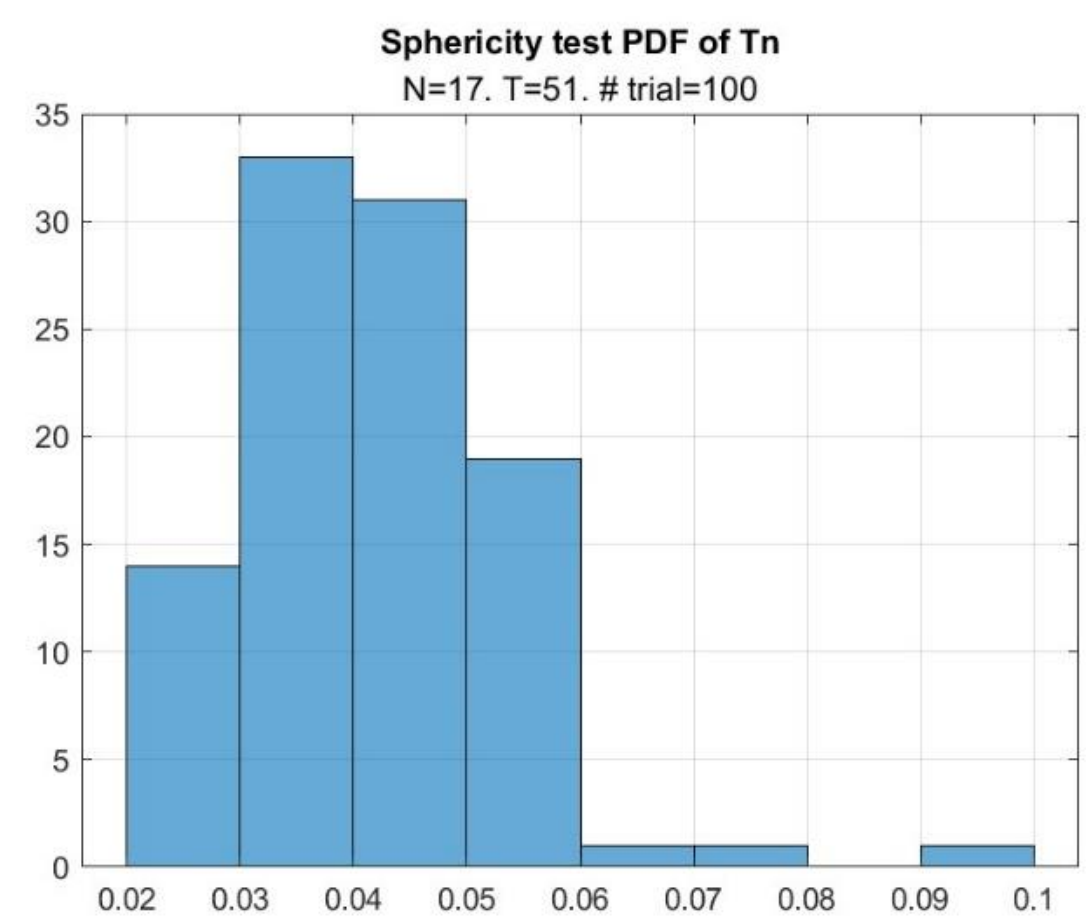

Fig. 33. Normal sphericity test PDF of $\boldsymbol{T}_N$. N=17. T=51.

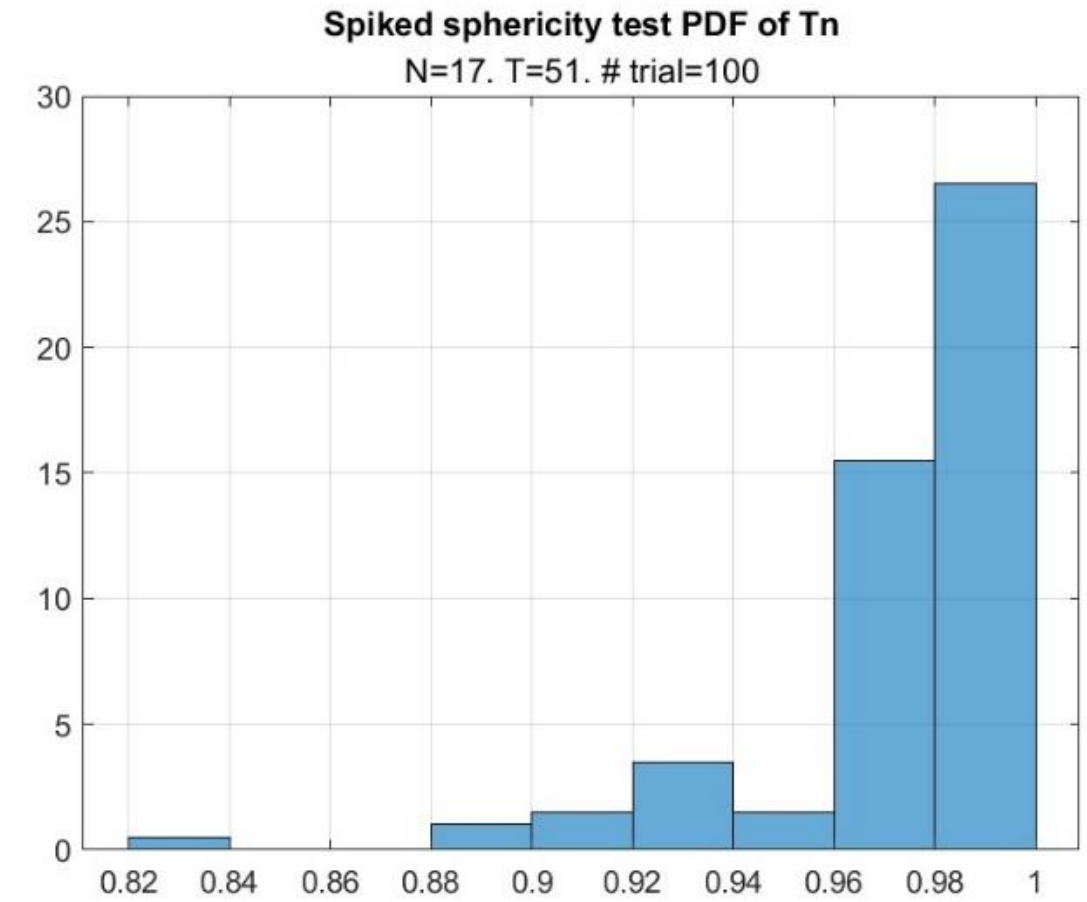

Fig. 34. Spiked sphericity test PDF of $\boldsymbol{T}_N$. N=17. T=51.

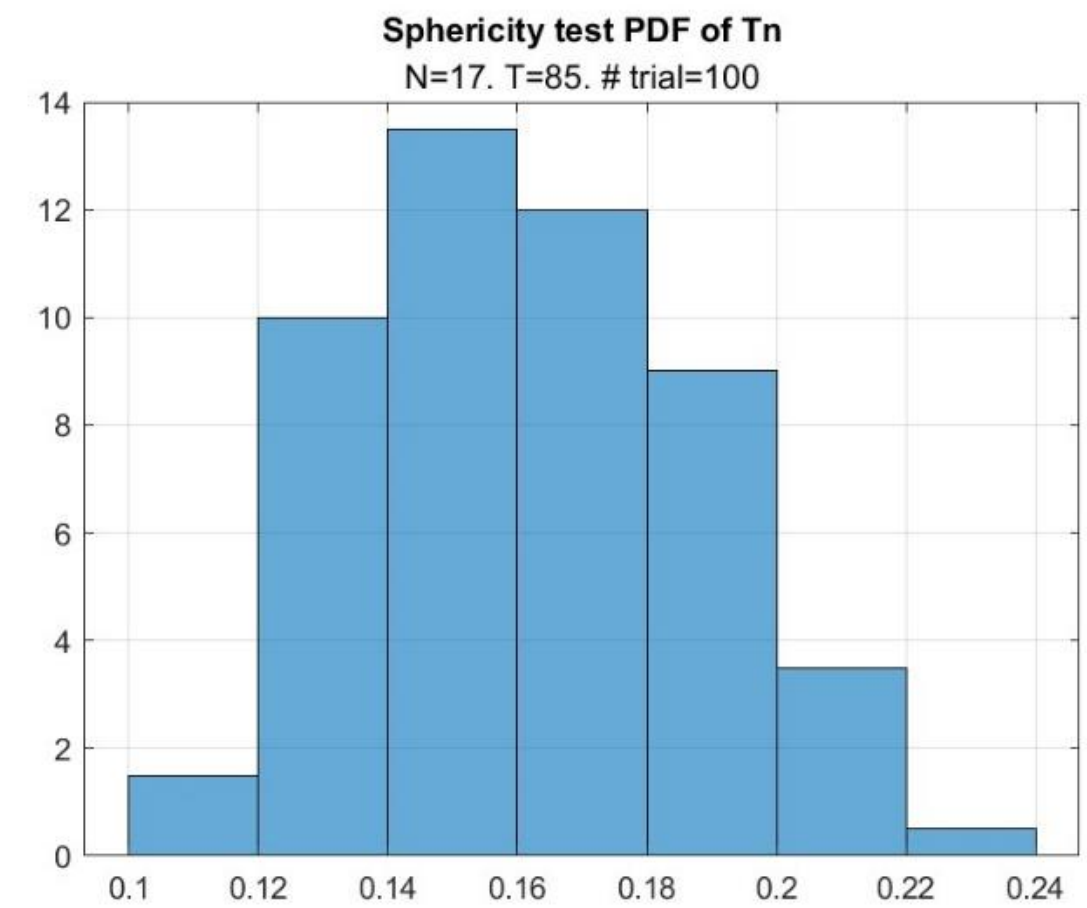

Fig. 35. Normal sphericity test PDF of $\boldsymbol{T}_N$. N=17. T=85.

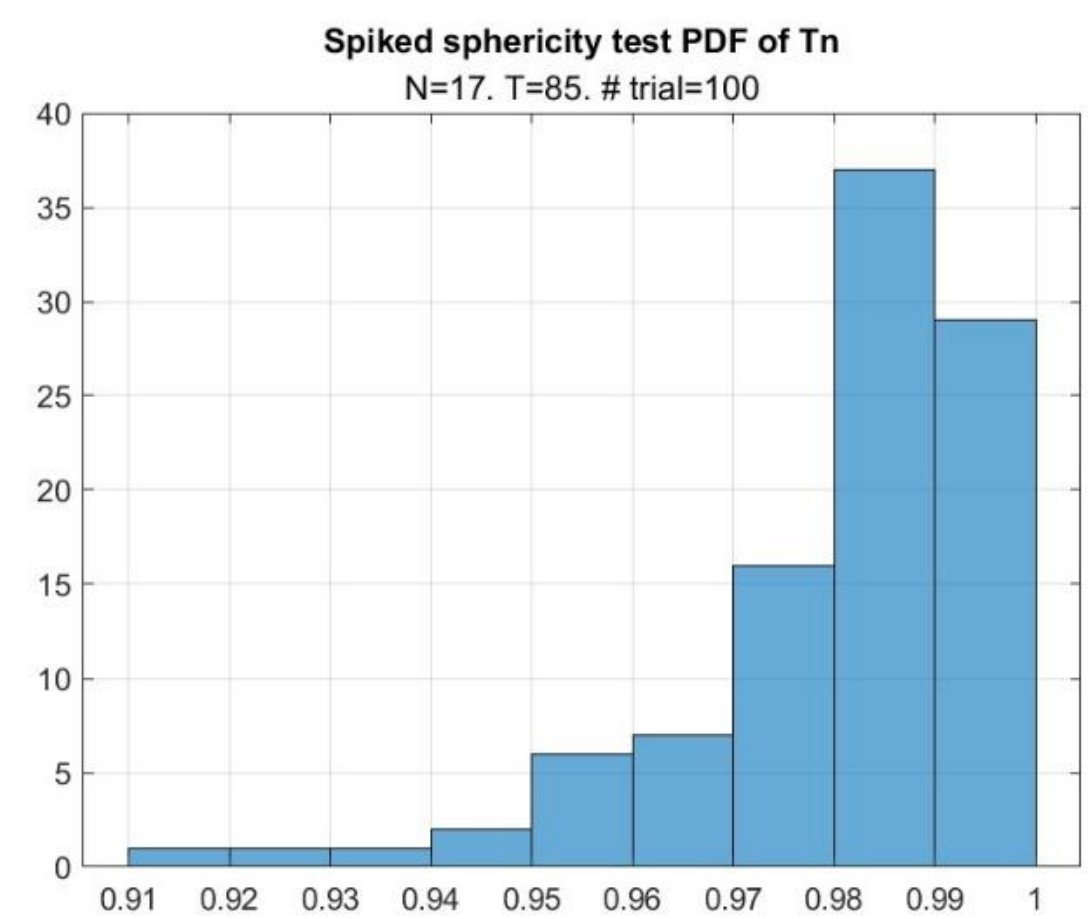

Fig. 36. Spiked sphericity test PDF of $\boldsymbol{T}_N$. N=17. T=85.

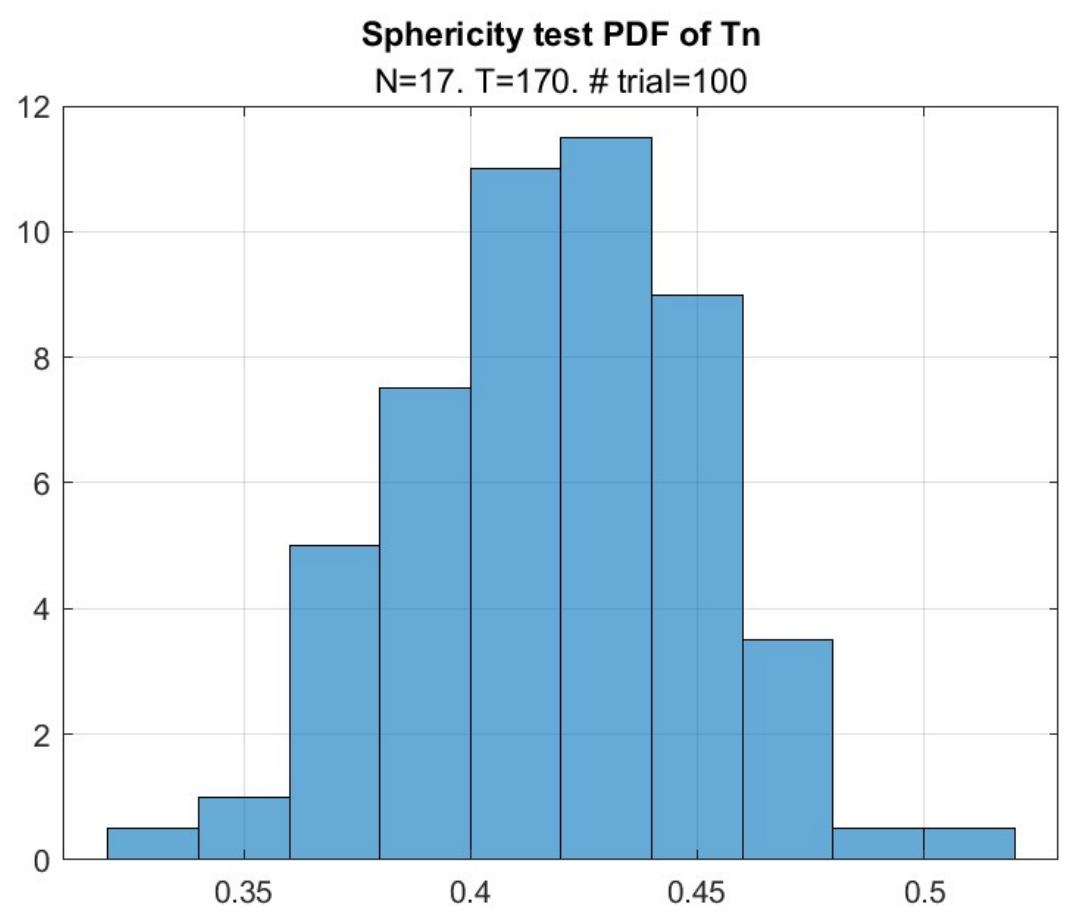

Fig. 37. Normal sphericity test PDF of $\boldsymbol{T}_N$. N=17. T=170.

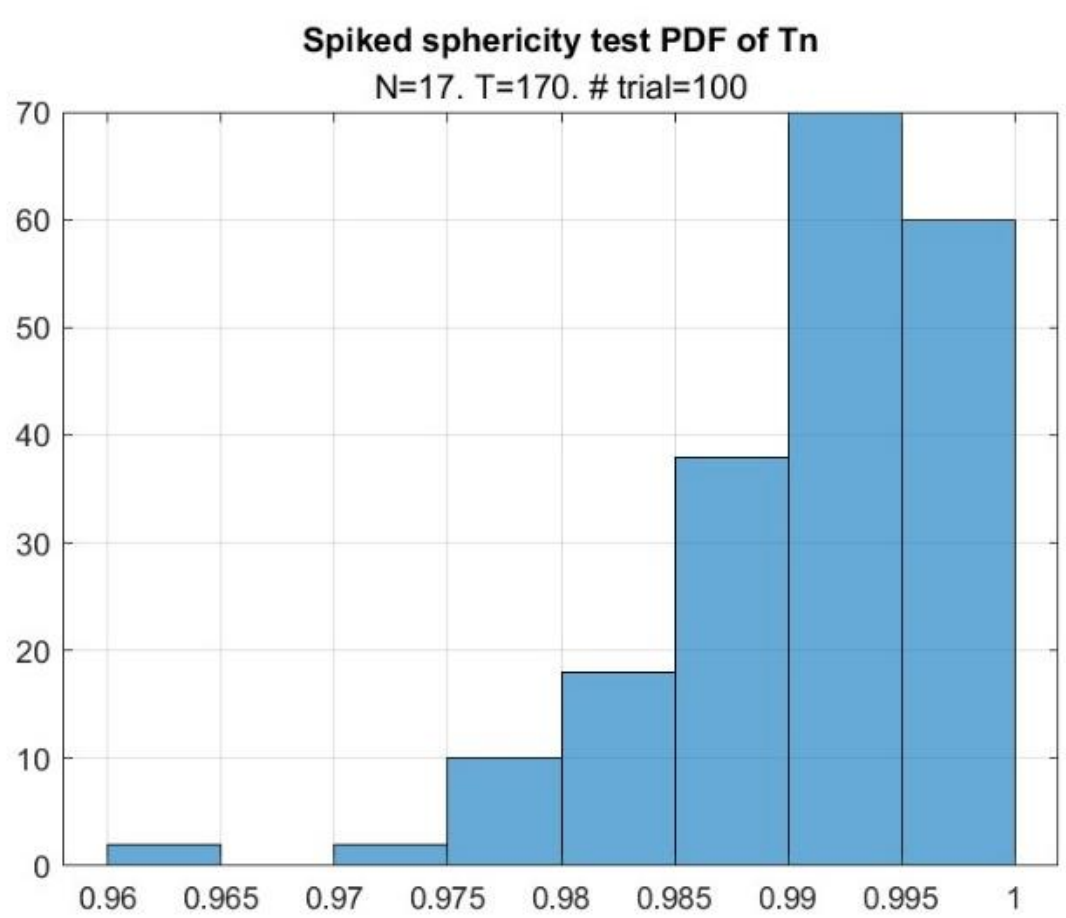

Fig. 38. Spiked sphericity test PDF of $\boldsymbol{T}_N$. N=17. T=170.

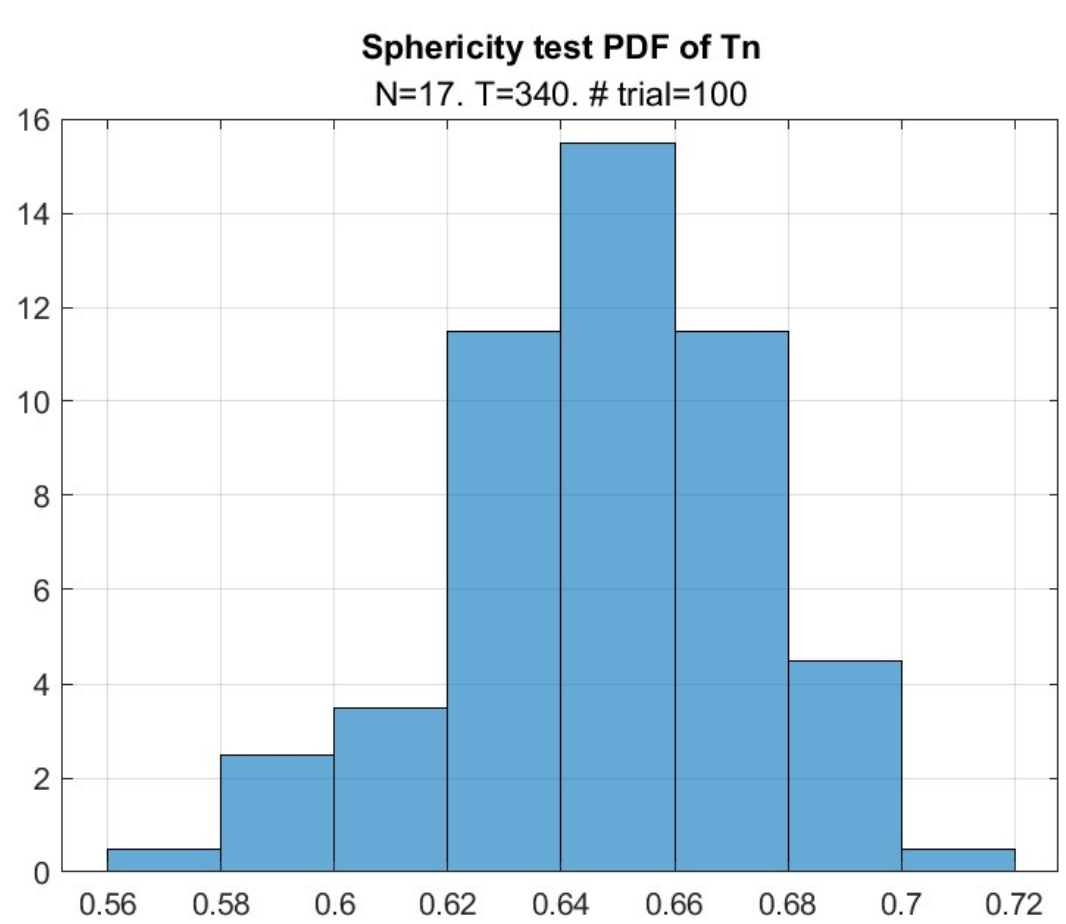

Fig. 39. Normal sphericity test PDF of $\boldsymbol{T}_N$. N=17. T=340.

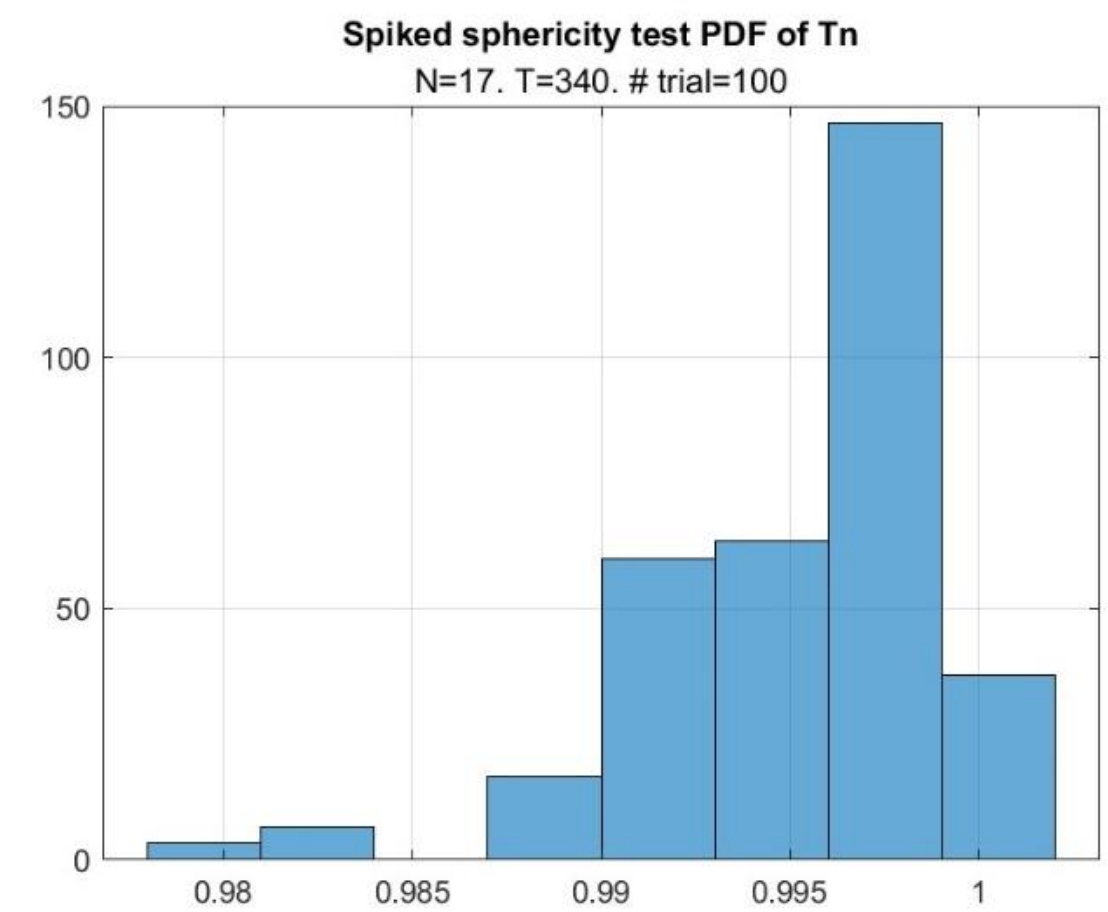

Fig. 40. Spiked sphericity test PDF of $\boldsymbol{T}_N$. N=17. T=340.

Analysis of this data leads to an important conclusion. Indeed, unlike redundancy averaging, the proposed transformation of the sample matrices into the p.d. Toeplitz Hermitian matrices gives a high likelihood of the transformed solutions. Indeed, the regular "sphericity" likelihood ratio remained well below $10^{-20}$ (!!!) over all 5000 iterations of the Newton's routine (being maximal ($10 \cdot 10^{-21}$) for the few initial iterations). Similarly, the relatively high "spiked" sphericity test for the initial solution fluctuated, not exceeding 0.6 for the rest of "rectification".

For the direct conversion with the same sample volume $T$ = 85, the regular sphericity test got the median value 0.015, with the median LR value = 0.15 for the true covariance matrix $\mathbf{T}_N$ (Fig. 27, Fig. 35). The "spiked" sphericity test reached the median value 0.97 (Fig. 28), while for the true matrix $\mathbf{T}_N$ the median value is 0.98 (Fig. 36). For the greater sample volume $T$ = 170 (Fig. 29) and $T$ = 340 (Fig. 31), the difference in the sphericity test values produced by the reconstructed p.d. Toeplitz matrix and the true covariance matrix $\mathbf{T}_N$ gets even smaller, as demonstrated by Fig. 37 and Fig. 39. Yet, the difference between these likelihood ratio values and the likelihood ratio values generated by the true covariance matrix $\mathbf{T}_N$ suggest that the derived Toeplitz matrices are not the ML estimates, and the solutions with the improved LR values may be still searched for.

One of the possibilities to improve the LR of the derived solutions could be "rectification" of the sample matrix $\widehat{\mathbf{R}}_N$ by replacing its eigenspectrum by the RMT methodology-advised eigenspectrum. Indeed, the RMT eigenvalues are closer to the actual eigenvalues of the matrix, which provides the hope that transformation of the sample matrix with the improved eigenvalues should lead to the "more likely" Toeplitz matrix.

In TABLE IV. for a particular $\widehat{\mathbf{R}}_N$ we provide its eigenvalues, along with the eigenvalues of the true matrix, and then report on the eigenvalues of the reconstructed Toeplitz matrix and the LR value produced by the solution. Then for the number of "noise subspace" eigenvalues, varying from $m$ = 2 to $m$ = 7, we report on the RMT-modified eigenvalues, the eigenvalues of the reconstructed p.d. Toeplitz matrices, and their LR values. For comparison, we replaced sample eigenvalues in the matrix $\widehat{\mathbf{R}}_N$ by the true (!) eigenvalues of the matrix $\mathbf{T}_N$, and reconstructed the p.d. Toeplitz matrix using the sample matrix with the true eigenvalues. This leads to unexpected results. In TABLE IV. ,

- $\mathbf{T}_N^{(0)}$ = reconstruction from $\widehat{\mathbf{R}}_N$
- $\mathbf{T}_N^{(1)}$ = reconstruction from $\widehat{\mathbf{R}}_N$ with its eigenvalues replaced by RMT-modified eigenvalues
- $\mathbf{T}_N^{(2)}$ = reconstruction from $\widehat{\mathbf{R}}_N$ with its eigenvalues replaced by $\mathbf{T}_N$ eigenvalues
- mod = eigenvalues normalized and multiplied by the sum of $\lambda^*$

TABLE IV.

| | |
|---|---|
| ***Sphericity test*** $T_N$ | 0.1494 |
| ***Sphericity test*** $T_N^{(0)}$ | 0.019860 |
| ***Sphericity test*** $T_N^{(2)}$ | 0.021491 |

| ***Multiplicity*** | ***Sphericity test*** $T_N^{(1)}$ |
|---|---|
| 2 | 0.018103 |
| 3 | 0.017656 |
| 4 | 0.009460 |
| 5 | 0.011467 |
| 6 | 0.001358 |
| 7 | 0.000002 |

| $\mathbf{T}_N^{(2)}$ | | | |
|---|---|---|---|
| $\lambda^*$ | $\lambda(\mathrm{T}_N)$ ***mod*** | $\lambda(T_N^{(0)})$ ***mod*** | $\lambda(T_N^{(2)})$ ***mod*** |
| 1.496411 | 1.496411 | 1.955747 | 1.756152 |
| 1.424830 | 1.424830 | 1.626097 | 1.495991 |
| 1.136760 | 1.136760 | 1.336983 | 1.412525 |
| 1.000872 | 1.000872 | 1.240047 | 1.204820 |
| 1.000083 | 1.000083 | 0.768082 | 0.797628 |
| 0.992370 | 0.992370 | 0.582221 | 0.622079 |
| 0.814989 | 0.814989 | 0.527602 | 0.596449 |
| 0.450598 | 0.450598 | 0.300269 | 0.399659 |
| 0.158410 | 0.158410 | 0.141783 | 0.192391 |
| 0.023622 | 0.023622 | 0.020964 | 0.022082 |
| 0.002041 | 0.002041 | 0.001205 | 0.001222 |
| 0.000210 | 0.000210 | 0.000320 | 0.000329 |
| 0.000104 | 0.000104 | 0.000128 | 0.000116 |
| 0.000100 | 0.000100 | 0.000099 | 0.000080 |
| 0.000100 | 0.000100 | 0.000067 | 0.000076 |
| 0.000100 | 0.000100 | 0.000045 | 0.000052 |
| 0.000100 | 0.000100 | 0.000041 | 0.000050 |

| **Multiplicity = 2** | | | |
|---|---|---|---|
| $\lambda^*$ | $\lambda(T_N)$ ***mod*** | $\lambda(T_N^{(0)})$ ***mod*** | $\lambda(T_N^{(1)})$ ***mod*** |
| 1.575354 | 1.597275 | 2.087572 | 2.024415 |
| 1.484907 | 1.520869 | 1.735702 | 1.822202 |
| 1.342875 | 1.213382 | 1.427101 | 1.383695 |
| 1.257612 | 1.068335 | 1.323631 | 1.304199 |
| 1.010768 | 1.067493 | 0.819854 | 0.808478 |
| 0.997429 | 1.059260 | 0.621465 | 0.631049 |
| 0.803576 | 0.869923 | 0.563165 | 0.582041 |
| 0.434592 | 0.480970 | 0.320508 | 0.332381 |
| 0.142716 | 0.169087 | 0.151340 | 0.161577 |
| 0.022383 | 0.025214 | 0.022377 | 0.022629 |
| 0.001851 | 0.002179 | 0.001286 | 0.001287 |
| 0.000235 | 0.000224 | 0.000341 | 0.000392 |
| 0.000101 | 0.000111 | 0.000137 | 0.000135 |
| 0.000095 | 0.000107 | 0.000105 | 0.000094 |
| 0.000094 | 0.000107 | 0.000072 | 0.000080 |
| 0.000081 | 0.000107 | 0.000048 | 0.000049 |
| 0.000081 | 0.000107 | 0.000044 | 0.000046 |

| **Multiplicity = 3** | | | |
|---|---|---|---|
| $\lambda^*$ | $\lambda(T_N)$ ***mod*** | $\lambda(T_N^{(0)})$ ***mod*** | $\lambda(T_N^{(1)})$ ***mod*** |
| 1.575354 | 1.597275 | 2.087572 | 2.012168 |
| 1.484907 | 1.520869 | 1.735702 | 1.787841 |
| 1.342875 | 1.213382 | 1.427101 | 1.380484 |
| 1.257612 | 1.068335 | 1.323631 | 1.311123 |
| 1.010768 | 1.067493 | 0.819854 | 0.813476 |
| 0.997429 | 1.059260 | 0.621465 | 0.638532 |
| 0.803576 | 0.869923 | 0.563165 | 0.592890 |
| 0.434592 | 0.480970 | 0.320508 | 0.344196 |
| 0.142716 | 0.169087 | 0.151340 | 0.168831 |
| 0.022383 | 0.025214 | 0.022377 | 0.023040 |
| 0.001851 | 0.002179 | 0.001286 | 0.001312 |
| 0.000235 | 0.000224 | 0.000341 | 0.000444 |
| 0.000101 | 0.000111 | 0.000137 | 0.000137 |
| 0.000095 | 0.000107 | 0.000105 | 0.000093 |
| 0.000085 | 0.000107 | 0.000072 | 0.000081 |
| 0.000085 | 0.000107 | 0.000048 | 0.000052 |
| 0.000085 | 0.000107 | 0.000044 | 0.000047 |

| **Multiplicity = 4** | | | |
|---|---|---|---|
| $\lambda^*$ | $\lambda(T_N)$ ***mod*** | $\lambda(T_N^{(0)})$ ***mod*** | $\lambda(T_N^{(1)})$ ***mod*** |
| 1.575354 | 1.597275 | 2.087572 | 2.024406 |
| 1.484907 | 1.520869 | 1.735702 | 1.778644 |
| 1.342875 | 1.213382 | 1.427101 | 1.368551 |

| Multiplicity = 4 | | | |
|---|---|---|---|
| $\lambda^*$ | $\lambda(T_N)$ *mod* | $\lambda(T_N^{(0)})$ *mod* | $\lambda(T_N^{(1)})$ *mod* |
| 1.257612 | 1.068335 | 1.323631 | 1.290264 |
| 1.010768 | 1.067493 | 0.819854 | 0.806205 |
| 0.997429 | 1.059260 | 0.621465 | 0.638583 |
| 0.803576 | 0.869923 | 0.563165 | 0.604426 |
| 0.434592 | 0.480970 | 0.320508 | 0.359487 |
| 0.142716 | 0.169087 | 0.151340 | 0.178035 |
| 0.022383 | 0.025214 | 0.022377 | 0.023479 |
| 0.001851 | 0.002179 | 0.001286 | 0.001470 |
| 0.000235 | 0.000224 | 0.000341 | 0.000801 |
| 0.000101 | 0.000111 | 0.000137 | 0.000136 |
| 0.000088 | 0.000107 | 0.000105 | 0.000086 |
| 0.000088 | 0.000107 | 0.000072 | 0.000078 |
| 0.000088 | 0.000107 | 0.000048 | 0.000052 |
| 0.000088 | 0.000107 | 0.000044 | 0.000046 |

| Multiplicity = 5 | | | |
|---|---|---|---|
| $\lambda^*$ | $\lambda(T_N)$ *mod* | $\lambda(T_N^{(0)})$ *mod* | $\lambda(T_N^{(1)})$ *mod* |
| 1.575354 | 1.597275 | 2.087572 | 2.009088 |
| 1.484907 | 1.520869 | 1.735702 | 1.772057 |
| 1.342875 | 1.213382 | 1.427101 | 1.369607 |
| 1.257612 | 1.068335 | 1.323631 | 1.295772 |
| 1.010768 | 1.067493 | 0.819854 | 0.810892 |
| 0.997429 | 1.059260 | 0.621465 | 0.640349 |
| 0.803576 | 0.869923 | 0.563165 | 0.606176 |
| 0.434592 | 0.480970 | 0.320508 | 0.363615 |
| 0.142716 | 0.169087 | 0.151340 | 0.181094 |
| 0.022383 | 0.025214 | 0.022377 | 0.023588 |
| 0.001851 | 0.002179 | 0.001286 | 0.001405 |
| 0.000235 | 0.000224 | 0.000341 | 0.000702 |
| 0.000090 | 0.000111 | 0.000137 | 0.000137 |
| 0.000090 | 0.000107 | 0.000105 | 0.000087 |
| 0.000090 | 0.000107 | 0.000072 | 0.000079 |
| 0.000090 | 0.000107 | 0.000048 | 0.000053 |
| 0.000090 | 0.000107 | 0.000044 | 0.000048 |

| Multiplicity = 6 | | | |
|---|---|---|---|
| $\lambda^*$ | $\lambda(T_N)$ *mod* | $\lambda(T_N^{(0)})$ *mod* | $\lambda(T_N^{(1)})$ *mod* |
| 1.575354 | 1.597275 | 2.087572 | 8.150122 |
| 1.484907 | 1.520869 | 1.735702 | 0.249964 |
| 1.342875 | 1.213382 | 1.427101 | 0.162773 |
| 1.257612 | 1.068335 | 1.323631 | 0.142314 |
| 1.010768 | 1.067493 | 0.819854 | 0.120227 |
| 0.997429 | 1.059260 | 0.621465 | 0.095497 |
| 0.803576 | 0.869923 | 0.563165 | 0.077282 |
| 0.434592 | 0.480970 | 0.320508 | 0.048816 |
| 0.142716 | 0.169087 | 0.151340 | 0.024938 |
| 0.022383 | 0.025214 | 0.022377 | 0.002593 |
| 0.001851 | 0.002179 | 0.001286 | 0.000151 |
| 0.000114 | 0.000224 | 0.000341 | 0.000014 |
| 0.000114 | 0.000111 | 0.000137 | 0.000013 |
| 0.000114 | 0.000107 | 0.000105 | 0.000012 |
| 0.000114 | 0.000107 | 0.000072 | 0.000012 |
| 0.000114 | 0.000107 | 0.000048 | 0.000012 |
| 0.000114 | 0.000107 | 0.000044 | 0.000010 |

| Multiplicity = 7 | | | |
|---|---|---|---|
| $\lambda^*$ | $\lambda(T_N)$ *mod* | $\lambda(T_N^{(0)})$ *mod* | $\lambda(T_N^{(1)})$ *mod* |
| 1.575354 | 1.597275 | 2.087572 | 4.007212 |
| 1.484907 | 1.520869 | 1.735702 | 1.794507 |
| 1.342875 | 1.213382 | 1.427101 | 1.225410 |
| 1.257612 | 1.068335 | 1.323631 | 0.868681 |
| 1.010768 | 1.067493 | 0.819854 | 0.530829 |
| 0.997429 | 1.059260 | 0.621465 | 0.260228 |
| 0.803576 | 0.869923 | 0.563165 | 0.228710 |
| 0.434592 | 0.480970 | 0.320508 | 0.122050 |
| 0.142716 | 0.169087 | 0.151340 | 0.028273 |
| 0.022383 | 0.025214 | 0.022377 | 0.006829 |
| 0.000363 | 0.002179 | 0.001286 | 0.000350 |
| 0.000363 | 0.000224 | 0.000341 | 0.000288 |
| 0.000363 | 0.000111 | 0.000137 | 0.000286 |
| 0.000363 | 0.000107 | 0.000105 | 0.000282 |
| 0.000363 | 0.000107 | 0.000072 | 0.000279 |
| 0.000363 | 0.000107 | 0.000048 | 0.000273 |
| 0.000363 | 0.000107 | 0.000044 | 0.000262 |

While the RMT-modified eigenvalues for the correct number of "noise subspace" eigenvalues are closer to the true eigenvalues of the covariance matrix $\mathbf{T}_N$, the reconstructed p.d. Toeplitz matrix demonstrates a much worse LR value than the matrix reconstructed from the unmodified sample matrix. In fact, only for the set of true eigenvalues that replaced the sample eigenvalues in the sample matrix $\widehat{\mathbf{R}}_N$, we got the p.d. reconstructed Toeplitz matrix with the LR value, slightly exceeding the LR value of the p.d. reconstructed Toeplitz matrix originated by the unmodified sample matrix. In all other cases, the replacement of the sample matrix eigenvalues by the “more accurate” RMT-derived ones only degraded the LR value produced by the reconstructed p.d. Toeplitz matrix. The exhaustive theoretical explanation of this phenomenon is not evident at the moment. The problem may be in the different axioms of the conventional ($N = \text{const}, T \to \infty$) and Kolmogorov's ($N \to \infty,\ T \to \infty,\ T/N \to c$) asymptotics, supporting the Maximum Likelihood and RMT consistency. Yet, if we still reside within the maximum likelihood paradigm, we have to admit that the LR improvement beyond provided by the proposed transformation of the unstructured ML (sample) matrix remains an open problem, if we ignore the brute force LR optimization option.

## V. Conclusions and Recommendations

The renewed interest in the problem of the Toeplitz covariance matrix estimation via the redundancy averaging (spatial averaging, Toeplitzification) algorithm has been ignited by the recent results that proved the asymptotic almost sure consistency of this algorithm under the Kolmogorov asymptotic conditions when the matrix dimension $N$ and the sample volume $T$ both tend to infinity ($N \to \infty,\ T \to \infty,\ T/N \to c$). In this study we first demonstrated that the proven asymptotic consistency in the spectral norm does not contradict with the generation by the redundancy averaging of a number of negative eigenvalues for quite typical covariance matrices. Moreover, we demonstrated that the sample volume, required for no negative eigenvalues to be generated, is proportional to $N/\lambda_{\min}^2$, which for $\lambda_{\min} \ll 1$ means an impractically large sample volume required. We demonstrated that the proper diagonal loading of the redundancy averaged matrix that mitigates the negative eigenvalues, results in an extremely low likelihood of the loaded matrices, tens of order of magnitude smaller than the likelihood of the true matrix, and therefore much smaller than the likelihood of the maximum likelihood solution.

We demonstrated that the attempts to rectify the redundancy averaged matrices by improving their eigenspectra to meet the suggested spectra by the Random Matrix Theory methodology [11] do not improve the likelihood of the "rectified" solutions with the improved eigenspectra. This phenomenon is closely related to the established non-uniqueness of the Hermitian Toeplitz matrices with the specified eigenspectrum. Therefore, the pursuit of the "proper" eigenvalues by the proposed iterative techniques, while leading to some improvement of the optimized eigenspectrum, does not lead to the likelihood growths. For that reason, in this study we explored an alternative technique of transition from the sample Hermitian matrix to the p.d. Toeplitz Hermitian one, proposed by us in [24]. We demonstrated that this method which reconstructs the Toeplitz matrix with the same Maximum Entropy spectrum as the sample matrix ME spectrum, always (!) results in p.d. Toeplitz Hermitian matrices with the LR values overwhelmingly greater than the LR values demonstrated by the properly loaded redundancy averaged covariance matrix.

To conclude our study, we have to admit that despite the recently proven (6) asymptotic almost sure convergence, the spatial averaging remains the simplest but one of the least efficient procedures for the transition from the unstructured maximum likelihood (sample) covariance matrix estimate to the ML-proxy p.d. Toeplitz Hermitian covariance matrix estimate. The transition from the sample matrix to the p.d. Hermitian Toeplitz matrix, proposed in [24], provides solutions with much higher likelihood values. At the same time, the attempts to rectify the redundancy averaged solutions by means of the Newton's iterative techniques that modify the eigenspectrum of the redundancy averaged covariance matrices were not very successful. The main reason is that the Hermitian Toeplitz matrices are not uniquely specified by the eigenspectrum, in which rectification leads to the "unlikely" solutions, though with the close to the desired eigenspectrum.

Finally, we demonstrated that the attempts to "improve" the sample matrix by replacement of its eigenvalues by the more accurate RMT-produced ones, contrary to expectations, leads to significant LR degradation of the reconstructed p.d. Toeplitz matrix. Only replacement of sample eigenvalues by the true eigenvalues of the covariance Toeplitz matrix led to the very insignificant improvement of the LR value, generated by the reconstructed p.d. Toeplitz matrix. Detailed theoretical explanation of this phenomenon is yet to come, while the main result of this study is the demonstration of much higher efficiency of the transformation, proposed in [24], of the sample matrix into a p.d. Toeplitz Hermitian covariance matrix estimate.